\documentclass[%
 superscriptaddress,
 aip,
 amsmath,amssymb,
reprint,%
 author-year,%
longbibliography
]{revtex4-2}

\usepackage[utf8]{inputenc}
\usepackage{
amsmath,
amssymb,
bm,
color,
comment,
dcolumn,
enumerate,
float,
graphicx,
hyperref,
mathrsfs,
tabularx,
titlesec,
siunitx,
soul,
subfigure,
multirow,
hhline
}

\DeclareSIUnit\molar{\mole\per\cubic\deci\metre}
\DeclareSIUnit\Molar{M}

\definecolor{linkColor}{rgb}{0.7,0,0}
\definecolor{darkred}{rgb}{0.7,0,0}
\hypersetup{pdfborder={0 0 0},colorlinks=true,urlcolor=linkColor,citecolor=linkColor}

\newcommand{\ii}{{i\mkern1mu}}
\newcommand{\ft}{\mathcal{F}}
\newcommand{\ift}{\mathcal{F}^{-1}}
\newcommand{\pyFDSTEM}{\href{https://github.com/py4dstem/py4DSTEM}{\texttt{py4DSTEM}}}
\newcommand{\prismatic}{\href{https://prism-em.com/}{\texttt{Prismatic}}}

\titleformat{\section}
{\color{darkred}\sffamily\bfseries}
{\color{darkred}\thesection}{1em}{}
\titleformat{\subsection}
{\color{darkred}\sffamily\itshape}
{\color{darkred}\thesection}{1em}{}
\titlespacing*{\section}
{0pt}{8pt}{0pt}
\titlespacing*{\subsection}
{0pt}{8pt}{0pt}

\emergencystretch=\maxdimen
\begin{document}

% \title[PRISM 2.0]{An Even Faster Algorithm for Scanning Transmission Electron Microscopy (STEM) Imaging and Diffraction Simulations}

\title[ACOM Correlation]{Automated Crystal Orientation Mapping in py4DSTEM using Sparse Correlation Matching}

% \title[ACOM Part I - Correlation]{Automated Crystal Orientation Mapping in py4DSTEM Part I - Sparse Correlation Matching}
% given Andrew's comments in the email, perhaps "Automated Crystal Orientation Mapping in py4DSTEM: (by/via/using)Sparse Correlation Matching (approach)" and the subsequent ML paper could be called ~"Automated Crystal Orientation Mapping in py4DSTEM: (by/via/using) Machine Learning (approach)". Gives the Part I part II feel without explicitly saying so.

\author{Colin Ophus}
\email{cophus@gmail.com}
\affiliation{National Center for Electron Microscopy, Molecular Foundry, Lawrence Berkeley National Laboratory, 1 Cyclotron Road, Berkeley, CA, USA, 94720}

\author{Steven E Zeltmann}
\affiliation{Department of Materials Science and Engineering, University of California, Berkeley, CA,  94720}

\author{Alexandra Bruefach}
\affiliation{Department of Materials Science and Engineering, University of California, Berkeley, CA,  94720}

\author{Alexander Rakowski}
\affiliation{National Center for Electron Microscopy, Molecular Foundry, Lawrence Berkeley National Laboratory, 1 Cyclotron Road, Berkeley, CA, USA, 94720}

\author{Benjamin H Savitzky}
\affiliation{National Center for Electron Microscopy, Molecular Foundry, Lawrence Berkeley National Laboratory, 1 Cyclotron Road, Berkeley, CA, USA, 94720}

\author{Andrew M Minor}
\affiliation{National Center for Electron Microscopy, Molecular Foundry, Lawrence Berkeley National Laboratory, 1 Cyclotron Road, Berkeley, CA, USA, 94720}
\affiliation{Department of Materials Science and Engineering, University of California, Berkeley, CA,  94720}

\author{MC Scott}
\affiliation{National Center for Electron Microscopy, Molecular Foundry, Lawrence Berkeley National Laboratory, 1 Cyclotron Road, Berkeley, CA, USA, 94720}
\affiliation{Department of Materials Science and Engineering, University of California, Berkeley, CA,  94720}

\date{\today}
\begin{abstract}

% iven that your sample can still only be one xstal thick, I am not sure that polycrystalline materials is the main driver. I think the main driver is the automation. I would suggest changing this to :"Crystalline materials for technical applications are often complex assemblies of different phases and oriented grains, often with incomplete information based on the diffracting condition probed by an electron beam.

% often contain multiple phases and orientations, with non-ideal diffracting conditions.

%  Robust  identification  of  the  phases  and  orientation  relationships from these samples is crucial,  but the information extracted from electron diffraction experiments is often incomplete.  
Crystalline materials used in technological applications are often complex assemblies composed of multiple phases and differently oriented grains. Robust identification of the phases and orientation relationships from these samples is crucial, but the information extracted from the diffraction condition probed by an electron beam is often incomplete. We therefore have developed an automated crystal orientation mapping (ACOM) procedure which uses a converged electron probe to collect diffraction patterns from multiple locations across a complex sample. We provide an algorithm to determine the orientation of each diffraction pattern based on a fast sparse correlation method. We test the speed and accuracy of our method by indexing diffraction patterns generated using both kinematical and dynamical simulations. We have also measured orientation maps from an experimental dataset consisting of a complex polycrystalline twisted helical AuAgPd nanowire. From these maps we identify twin planes between adjacent grains, which may be responsible for the twisted helical structure. All of our methods are made freely available as open source code, including tutorials which can be adapted to perform ACOM measurements on diffraction pattern datasets. 

\vspace{1 cm}

\end{abstract}
% \pacs{PACS Numbers}
% \keywords{Keywords go here}
\maketitle

\section*{Introduction}

Polycrystalline materials are ubiquitous in technological applications. An ideal crystal structure can be fully defined with a small number of parameters: the 3 vectors defining its unit cell, and the position and species of each atom inside the unit cell \citep{borchardt2011crystallography}. To fully describe crystalline materials in the real world however, we require a description of both the crystal lattice, and all defects present in a given material. These include point defects such as dopants, vacancies, or interstitials \citep{dederichs1978lattice}, line defects such as dislocations \citep{lesar2014simulations}, planar defects including internal boundaries and surfaces \citep{tang2006diffuse}, and finally volume defects such as local strain fields \citep{janssen2007stress}. One large subset of crystalline materials are polycrystalline phases, which consist of many small crystalline grains, arranged in either a random or organized fashion. Many material properties such as mechanical strength \citep{thompson2000structure}, optical response \citep{park2019efficient, londono2021intrinsic}, or thermal or electrical conductivity \citep{castro2019role} are strongly modulated by the density and orientation of the boundaries between crystalline grains \cite{thompson1995texture}. Thus characterizing the orientation of polycrystalline grains is essential to understanding these materials.

The two primary tools used to study the orientation of polycrystalline materials are electron backscatter diffraction (EBSD) in scanning electron microscopy (SEM), and transmission electron microscopy (TEM). EBSD can measure the orientation of crystalline grains with very high accuracy, but has limited resolution and is primarily sensitive to the surface of materials \citep{humphreys2001review, wright2011review, wright2015introduction}. Alternatively, we can directly measure the atomic-scale structure and therefore the orientation of polycrystalline grains, either by using plane wave imaging in TEM \citep{li2020constrained}, or by focusing the probe down to sub-atomic dimensions and scanning over the sample surface in scanning TEM (STEM) \citep{peter2018segregation}. This is possible due to the widespread deployment of aberration correction for both TEM and STEM instruments \citep{linck2016chromatic, ramasse2017twenty}. Atomic resolution imaging, howeve,r strictly limits the achievable field-of-view, and requires relatively thin samples, and thus is primarily suited for measuring polycrystalline grain orientations of 2D materials \citep{ophus2015large, qi2020near}.

Another approach to orientation mapping in TEM is to use diffraction space measurements. For crystalline materials, diffraction patterns will contain Bragg spots with spacing inversely proportional to the spacing of atomic planes which are approximately perpendicular to the beam direction (described by both the Laue condition and Bragg equations \citep{fultz2012transmission}). To generate a spatially-resolved orientation map, we can focus a STEM probe down to dimensions of 0.5 to 50 nm, scan it over the sample surface, and record the diffraction pattern for each probe position, a technique referred to as nanobeam electron diffraction (NBED) \citep{ozdol2015strain}, scanning electron nanobeam diffraction (SEND) \citep{tao2009direct}, or four dimensional-scanning transmission electron microscopy (4D-STEM) (we choose this nomenclature for this text)  due to the 4D shape of the collected data \citep{bustillo20214d}. 4D-STEM experiments are increasingly enabled by fast direct electron detectors, as these cameras allow for much faster recording and much larger fields of view \citep{ophus2019four, nord2020fast, paterson2020fast}. 

By performing template matching of diffraction pattern libraries on 4D-STEM datasets, we can map the orientation of all crystalline grains with sufficient diffraction signal. This method is usually named automated crystal orientation mapping (ACOM), and has been used by many authors in materials science studies \citep{zaefferer1994line, rauch2005rapid, kobler2013combination, maclaren2020comparison, londono20201d, jeong2021automated, zuo2021strategies}. ACOM experiments in 4D-STEM are highly flexible; two recent examples include \cite{lang2021automated} implementing ACOM measurements in liquid cell experiments, and  \cite{wu2021seeing} adapting the ACOM method to a scanning confocal electron diffraction (SCED) experimental configuration. ACOM is also routinely combined with precession electron diffraction, where the STEM beam is continually rotated around a cone incident onto the sample, in order to excite more diffraction spots and thus produce more interpretable diffraction patterns \citep{brunetti2011confirmation, moeck2011high, eggeman2015scanning}. Recently, \cite{mehta2020unravelling} have combined simulations with machine learning segmentation to map orientations of 2D materials, and \cite{yuan2021training} have used machine learning methods to improve the resolution and sensitivity of orientation maps by training on simulated data.

In this study, we introduce a new sparse correlation framework for fast calculation of orientation maps from 4D-STEM datasets. Our method is based on template matching of diffraction patterns along only the populated radial bands of a reference crystal's reciprocal lattice, and uses direct sampling of the first two Euler angles, and a fast Fourier transform correlation step to solve for the final Euler angle. We test our method on both kinematical calculations, and simulated diffraction experiments incorporating dynamical diffraction. Finally, we generate orientation maps of polycrystalline AuAgPd helically twisted nanowires, and use clustering to segment the polycrystalline structure, and map the shared (111) twin planes of adjacent grains.% Future parts of this study will improve our approach using machine learning methods \cite{munshi2021ml}, and will extend our ACOM methods to multibeam electron diffraction experiments \citep{hong2021multibeam}.

% \section*{Theory}
\section*{Methods}

\begin{figure*}[htbp]
    \centering
    \includegraphics[width=6.4
    in]{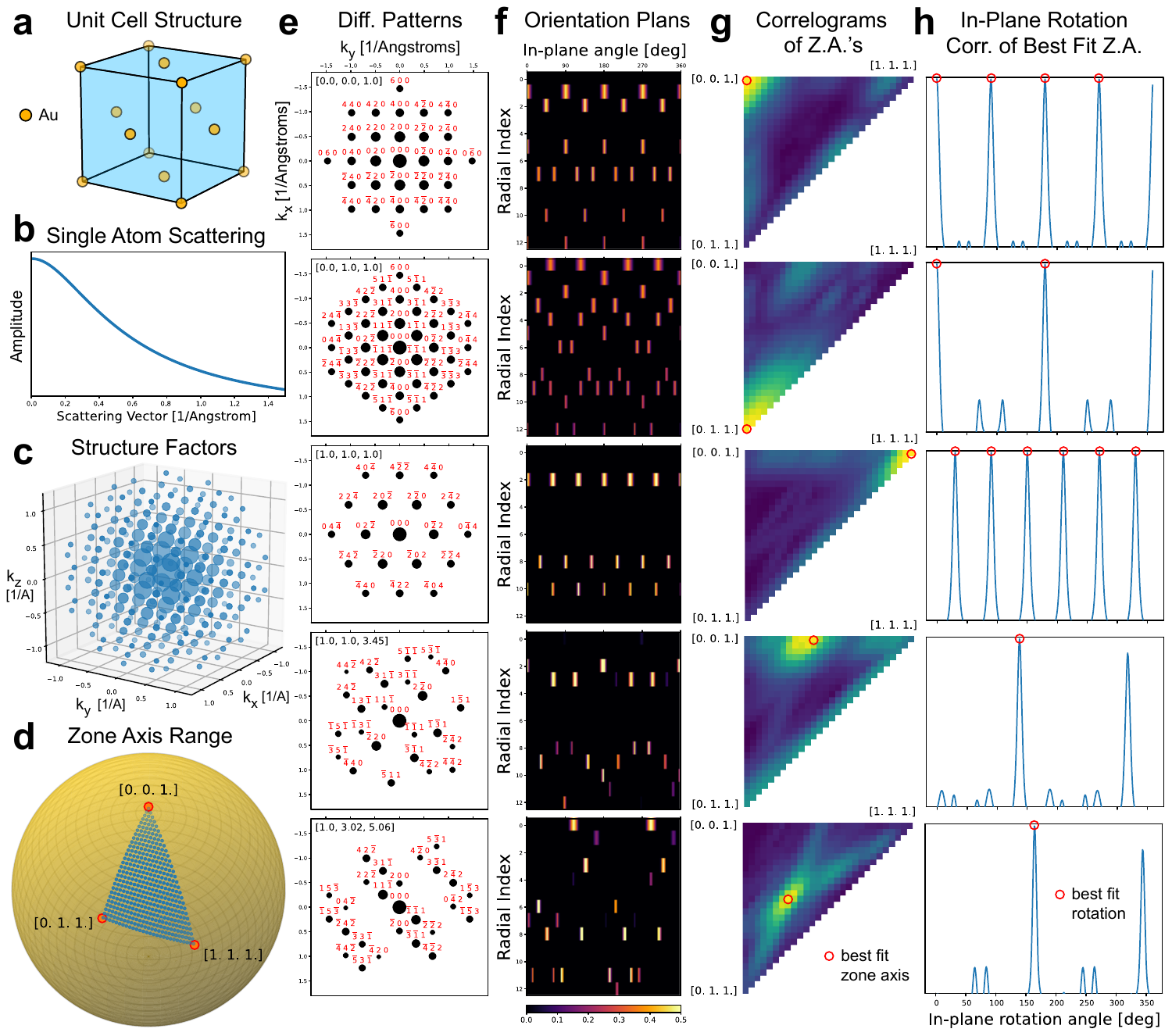}
    \caption{{\bf ACOM using correlation matching in \pyFDSTEM{}}. (a) Structure of fcc Au. (b) Atomic scattering factor of Au. (c) Structure factors for fcc Au. (d) Zone axes included in orientation plan. (e) Diffraction patterns for various orientations, and (f) corresponding orientation plan slices. (g) Correlogram maxima for each pattern in (e) as a function of zone axis, and (h) corresponding in-plane rotation correlation. Highest correlation scores are shown in (g) and (h) using red circles.}
    \label{Fig:ACOM_corr_match}
\end{figure*}

\subsection*{Structure Factor Calculations}

The \emph{structure factors} of a given crystalline material are defined as the complex coefficients of the Fourier transform of an infinite crystal \citep{spence1993accurate}. We require these coefficients in order to simulate kinematical diffraction patterns, and thus we briefly outline their calculation procedure here.

First, we define the reference crystal structure. This structure consists of two components, the first being its unit cell defined by its lattice vectors $\bm{a}$, $\bm{b}$, and $\bm{c}$ composed of positions in $\bm{r}=(x,y,z)$, the 3D real space coordinate system. The second component of a crystal structure is an array with dimensions $[N, 4]$ containing the fractional atomic positions $\bm{p}_n = (p_{\bm{a}}, p_{\bm{b}}, p_{\bm{c}})_n$ and atomic number $Z_n$, for the $n$th index of $N$ total atoms in the unit cell. Together these positions and atomic numbers are referred to as the atomic basis. Because $\bm{p}_n$ is given in terms of the lattice vectors, all fractional positions have values inside the range $[0, 1)$. The unit cell and real space Cartesian coordinates of the fcc Au structure are plotted in Fig.~\ref{Fig:ACOM_corr_match}a.

All subsequent calculations are performed in reciprocal space (also known as Fourier space or diffraction space). Thus the next step is to compute the reciprocal lattice vectors, defined by \citep{gibbs1884elements}
\begin{eqnarray}
    \bm{a}^*
&=&
    \frac{\bm{b} \times \bm{c}}{
    \bm{a} \cdot [\bm{b} \times \bm{c}]}
    =
    \frac{\bm{b} \times \bm{c}}{V}
    \nonumber 
    \\
    \bm{b}^*
&=&
    \frac{\bm{c} \times \bm{a}}{
    \bm{b} \cdot [\bm{c} \times \bm{a}]}
    =
    \frac{\bm{c} \times \bm{a}}{V}
    \nonumber 
    \\
    \bm{c}^*
&=&
    \frac{\bm{a} \times \bm{b}}{
    \bm{c} \cdot [\bm{a} \times \bm{b}]}
    =
    \frac{\bm{a} \times \bm{b}}{V},
\end{eqnarray}
where $\times$ represents the vector cross product and $V$ is the cell volume in real space. Note that this definition does not include factors of $2\pi$, and therefore all reciprocal coordinates have spatial frequency units.

Next, we calculate the position of all reciprocal lattice points required for our kinematical diffraction calculation, given by
\begin{equation}
    \bm{g}_{hkl} 
    =
    h \, \bm{a}^* + 
    k \, \bm{b}^* + 
    l \, \bm{c}^*,
\end{equation}
where $h$, $k$, and $l$ are integers representing the reciprocal lattice index points corresponding to the Miller indices $(h,k,l)$. We include only points where $|\bm{q}_{hkl}| < q_{\rm{max}}$, where $\bm{q}=(q_x,q_y,q_z)$ are the 3D coordinates in reciprocal space, i.e.\ those which fall inside a sphere given by the maximum scattering vector $q_{\rm{max}}$. To find all reciprocal lattice coordinates, we first determine the shortest vector given by linear combinations of $(\bm{a}^*,\bm{b}^*,\bm{c}^*)$, and divide $q_{\rm{max}}$ by this vector length to give the range for $(h,k,l)$. We then tile $(h,k,l)$ in both the positive and negative directions up to this value, and then remove all points with vector lengths larger than $q_{\rm{max}}$.

The reciprocal lattice defined above represents all possible coordinates where the structure factor coefficients $V_g(\bm{q})$ could be non-zero. The structure factor coefficients depend only the atomic basis and are given by
\begin{equation}
    F_{hkl} = 
    \sum_{n=1}^N 
    f_n(|\bm{g}_{hkl}|)
    \exp\left[
    -2 \pi \ii (h,k,l) \cdot \bm{p}_n
    \right],
\end{equation}
where $f_n$ are the the single-atom scattering factors for the $n$th atom, which describe the scattering amplitude for a single atom isolated in space. There are multiple ways to parameterize $f_n$, but here we have chosen to use the factors defined by \cite{lobato2014accurate} which are implemented in \pyFDSTEM{}. Fig.~\ref{Fig:ACOM_corr_match}b shows the atomic scattering factor for an Au atom.

We have now defined all structure factor coefficients for a perfect infinite crystal as
\begin{equation}
  V_g(\bm{q}) =
  \begin{cases}
    F_{hkl} & \text{if $\bm{q} = \bm{g}_{hkl}$} 
    \\
    0 & \text{otherwise}.
  \end{cases}
\end{equation}
Fig.~\ref{Fig:ACOM_corr_match}c shows the structure factors of fcc Au, where the marker size denotes the intensity (magnitude squared) of the $F_{hkl}$ values.

% Many materials however have multiple atoms of the same species in their unit cell at high-symmetry points, which can lead zero-valued coefficients at some lattice coordinates. 

% The final 

\subsection*{Calculation of Kinematical Diffraction Patterns}

Here we briefly review the theory of kinematical diffraction of finite crystals, following \cite{de2003introduction}. We can fully describe an electron plane wave by its wavevector $\bm{k}$, which points in the direction of the electron beam and has a length given by $|\bm{k}| = 1/\lambda$, where $\lambda$ is the (relativistically-corrected) electron wavelength. Bragg diffraction of the electron wave along a direction $\bm{k}'$ occurs when electrons scatter from equally spaced planes in the crystal, described in reciprocal space as
\begin{equation}
    \bm{k}' = \bm{k} + \bm{g}_{hkl}.
    \label{eq:diffraction_condition}
\end{equation}
For elastic scattering, $\bm{k}'$ has the same length as $\bm{k}$, and so scattering can only occur along the spherical surface known as the \emph{Ewald sphere construction} \citep{ewald1921berechunung}. This expression will almost never be satisfied by a perfect infinite crystal. However, real samples have finite dimensions, and thus the Fourier transform of their lattice will include a \emph{shape factor} $D(\bm{q})$ convolved with each reciprocal lattice point. Thus diffraction can still occur, as long as Eq.~\ref{eq:diffraction_condition} is approximately satisfied.

If the sample foil is tilted an angle $\alpha$ away from the beam direction, the vector between a reciprocal lattice point $\bm{g}$ and its closest point on the Ewald sphere has a length equal to
\begin{equation}
    s_{\bm{g}} = 
    \frac{-\bm{g} \cdot(2 \bm{k} + \bm{g}) }
    {2 | \bm{k} + \bm{g} | \cos(\alpha) }.
    \label{eq:excitation error}
\end{equation}
The $s_{\bm{g}}$ term is known as the \emph{excitation error} of a given reciprocal lattice point $\bm{g}$. When the excitation error $s_{\bm{g}}=0$, the Bragg condition is exactly satisfied. When the length of $s_{\bm{g}}$ is on the same scale as the extent of the shape factor, the Bragg condition is approximately satisfied.

A typical TEM sample can be approximately described as a slab or foil which is infinite in two dimensions, and with some thickness $t$ along the normal direction. The shape function of such a sample is equal to
\begin{equation}
    D(q_z) = \frac{\sin(\pi q_z t)}{\pi q_z}.
    \label{Eq:shape_foil}
\end{equation}
Because this expression is convolved with each reciprocal lattice point, we can replace $q_z$ with the distance between the Ewald sphere and the reciprocal lattice point. For the orientation mapping application considered in this paper, we assume that $\alpha = 0$, and that the sample thickness $t$ is unknown. Instead, we replace Eq.~\ref{Eq:shape_foil} with the approximation
\begin{equation}
    D(q_z) = 
    \exp\left(
    -\frac{{q_z}^2}{2 \sigma^2}
    \right),
    \label{Eq:shape_gaussian}
\end{equation}
where $\sigma$ represents the excitation error tolerance for a given diffraction spot to be included. We chose this expression for the shape function because it decreases monotonically with increasing distance between the diffraction spot and the Ewald sphere $q_z$, and produces smooth output correlograms.

%\hl{Maybe note that we also chose this because it decreases monotonically with $s$?}

To calculate a kinematic diffraction pattern for a given orientation $\bm{w}$, we loop through all reciprocal lattice points and use Eq.~\ref{eq:excitation error} to calculate the excitation errors. The intensity of each diffraction spot is given by the intensity of the structure factor $|F_{hkl}|^2$, reduced by a factor defined by either Eq.~\ref{Eq:shape_foil} or Eq.~\ref{Eq:shape_gaussian}. We define the position of the diffraction spots in the imaging plane by finding two vectors perpendicular to the beam direction, and projecting the diffraction vectors $q$ into this plane. The result is the intensity of each spot $I_m$, and its two spatial coordinates $(q_{m_x},q_{m_y})$, or alternatively their polar coordinates $q_m = \sqrt{{q_{m_x}}^2 + {q_{m_y}}^2}$ and $\gamma_m =$ arctan2$(q_{m_y}, q_{m_x})$. Note that the in-plane rotation angle is arbitrarily defined for kinematical calculations in the forward direction. The resulting diffraction patterns are defined by the list of $M$ Bragg peaks $(q_{m_x},q_{m_y},I_m)$ or $(q_m,\gamma_m ,I_m)$.

Fig.~\ref{Fig:ACOM_corr_match}e shows diffraction patterns for fcc Au, along five different zone axes (orientation directions). Each pattern includes Bragg spots out to a maximum scattering angle of $q_{\rm{max}} = 1.5 \, \rm{\AA}^{-1}$, and each spot is labeled by the $(hkl)$ indices. The marker size shown for each spot scales with the amplitude of each spot's structure factor, decreased by Eq.~\ref{Eq:shape_gaussian} using $\sigma = 0.02 \, \rm{\AA}^{-1}$.

\subsection*{Generation of an Orientation Plan}

The problem we are solving is to identify the relative orientation between a given diffraction pattern measurement and a parent reference crystal. This orientation can be uniquely defined by a $[3 \times 3]$-size matrix $\tensor{\bm{m}}$, which rotates vectors $\bm{d}_0$ in the sample coordinate system to vectors $\bm{d}$ in the parent crystal coordinate system
% \begin{equation}
% \begin{bmatrix}
% 1 & 2 & 3\\
% a & b & c
% \end{bmatrix}
% \end{equation}
\begin{eqnarray}
    \begin{bmatrix}
       d_x \\
       d_y \\
       d_z
    \end{bmatrix}
    &=& 
    \left[
      \begin{array}{ccc}
        u_x & v_x & w_x \\
        u_y & v_y & w_y \\
        u_z & v_z & w_z  \\  
      \end{array}
    \right]
    \begin{bmatrix}
       d_{0_x} \\
       d_{0_y} \\
       d_{0_z}
    \end{bmatrix}
    \nonumber \\
    \bm{d} 
    &=& 
    \tensor{\bm{m}}
    \ \bm{d}_0,
\end{eqnarray}
where the first two columns of $\tensor{\bm{m}}$ given by $\bm{u}$ and $\bm{v}$ represent the orientation of the in-plane $x$ and $y$ axis directions of the parent crystal coordinate system, respectively, and the third column $\bm{w}$ defines the zone axis or out-plane-direction. The orientation matrix can be defined in many different ways, but we have chosen to use a $Z-X-Z$ Euler angle scheme \citep{rowenhorst2015consistent}, defined as
\begin{equation}
    \tensor{\bm{m}} =
    \left[
      \begin{array}{ccc}
        C_1 & -S_1 & 0 \\
        S_1 & C_1 & 0 \\
        0 & 0 & 1  \\  
      \end{array}
    \right]
    \left[
      \begin{array}{ccc}
        1 & 0 & 0  \\  
        0 & C_2 & S_2 \\
        0 & -S_2 & C_2 \\
      \end{array}
    \right]
    \left[
      \begin{array}{ccc}
        C_3 & -S_3 & 0 \\
        S_3 & C_3 & 0 \\
        0 & 0 & 1  \\   
      \end{array}
    \right],
    \label{Eq:rotation_matrices}
\end{equation}
where $C_1 = \cos(\phi_1)$, $S_1 = \sin(\phi_1)$, $C_2 = \cos(\theta_2)$, $S_2 = \sin(\theta_2)$, $C_3 = \cos(\phi_3)$, and  $S_3 = \sin(\phi_3)$. The Euler angles $(\phi_1, \theta_2, \phi_3)$ chosen are fairly arbitrarily, as are the signs of rotation matrices given above. This convention was chosen to preserve internal consistency and to produce orientation matrices with sorted directions.

In order to determine the orientation $\tensor{\bm{m}}$ of a given diffraction pattern, we use a two-step procedure. The first step is to calculate an \emph{orientation plan} $ P((\phi_1, \theta_2), \phi_3, q_s)$ for a given reference crystal. The second step, which is defined in the following section, is to generate a \emph{correlogram} from each reference crystal, from which we directly determine the correct orientation. 

The first two Euler angles $\phi_1$ and $\theta_2$ represent points on the unit sphere which will become the zone axis of a given orientation. The first step in generating an orientation plan is to select 3 vectors delimiting the extrema of the unique, symmetry-reduced zone axes possible for a given crystal. Fig.~\ref{Fig:ACOM_corr_match}d shows these boundary vectors for fcc Au, which are given by the directions $[001]$, $[011]$, and $[111]$. We next choose a sampling rate or angular step size, and generate a grid of zone axes to test. We define a 2D grid of vectors on the unit sphere which span the boundary vectors by using spherical linear interpolation (SLERP) formula defined by \cite{shoemake1985animating}. These points with a step size of $2^\circ$ are shown in Fig.~\ref{Fig:ACOM_corr_match}d. The rotation matrices which transform the zone axis vector (along the z axis) are given by the matrix inverse of the first two terms in Eq.~\ref{Eq:rotation_matrices}.

We then examine the vector lengths of all non-zero reciprocal lattice points $\bm{g}_{hkl}$ and find all unique spherical shell radii $q_s$. These radii will become the first dimension of our orientation correlogram, where each radius is assigned one index $s$. We loop through all included zone axes, and calculate a polar coordinate representation of the kinematical diffraction patterns. 

For each zone axis, the first step to compute the plan is to rotate all structure factor coordinates by the matrix inverse of the first two terms in Eq.~\ref{Eq:rotation_matrices}. Next, we compute the excitation errors $s_g$ for all peaks assuming a $[0,0,1]$ projection direction, and the in-plane rotation angle of all peaks $\gamma_{\bm{q}}$.  The intensity values of the orientation plan for all $q_s$ shells and in-plane rotation values $\phi_3$ are defined using the expression
% BACKUP:
% \begin{eqnarray}
%     && P((\phi_1, \theta_2), \phi_3, q_s) = 
%     A(\phi_1, \theta_2)
%     \sum_{\bm{g} \in q_s} 
%     q_s |F_{hkl}|
%     \nonumber \\
%     &&
%     \rm{max} \left\{
%         1 - \sqrt{
%         {s_g}^2 + 
%         \left[
%         \rm{mod}(
%         \phi_3 - \gamma_{\bm{q}} + \pi, 2 \pi) - \pi
%         \right]^2 {q_s}^2
%         }, 0
%     \right\},
%     \nonumber 
% \end{eqnarray}
\begin{eqnarray}
    && P_0((\phi_1, \theta_2), \phi_3, q_s) = 
    \sum_{ \left\{ \bm{g} \,:\, |\bm{g}| = q_s  \right\} } 
    {q_s}^\gamma |V_{\bm{g}}|^\omega \times
    \nonumber \\
    &&
    \rm{max} \left\{
        1 - \frac{1}{\delta} \sqrt{
        s_{\bm{g}}^2 + 
        \left[
        \rm{mod}(
        \phi_3 - \gamma_{\bm{g}} + \pi, 2 \pi) - \pi
        \right]^2 {q_s}^2
        }, 0
    \right\},
    \nonumber 
\end{eqnarray}
where $\delta$ is the correlation kernel size, $\gamma$ and $\omega$ represent the power law scaling for the radial and peak amplitude terms respectively, $\rm{max}(...)$ is the maximum function, which returns the maximum of its two arguments, $\rm{mod}(...)$ is the modulo operator, and the summation includes only those peaks $\bm{g}$  which belong to a given radial value $q_s$. We have used the combined indexing notation for $(\phi_1, \theta_2)$ to indicate that in practice, this dimension of the correlation plan contains all zone axes, and thus the entire array has only 3 dimensions. The correlation kernel size $\delta$ defines the azimuthal extent of the correlation signal for each reciprocal lattice point.  Note that Eqs.~\ref{Eq:shape_gaussian} and \ref{Eq:shape_foil} are not used for the calculation of orientation plans.

We normalize each zone axis projection using the function
\begin{equation}
    A(\phi_1, \theta_2) 
    = \frac{1}{\sqrt{
    \sum_{\phi_3}
    \sum_{q_s}
    {P_0((\phi_1, \theta_2), \phi_3, q_s)}^2
    }},
    \nonumber
\end{equation}
yielding the final normalized orientation plan
\begin{equation}
    P((\phi_1, \theta_2), \phi_3, q_s) = A(\phi_1, \theta_2) P_0((\phi_1, \theta_2), \phi_3, q_s)
\end{equation}
By default, we have weighted each term in the orientation plan with the prefactor $q_s |V_g|$, i.e.\ setting $\gamma=\omega=1$. The $q_s$ term gives slightly more weight to higher scattering angles, while the $|V_g|$ term is used to weight the correlation in favour of peaks with higher structure factor amplitudes, which was found to be more reliable than weighting the orientation plan by $|V_g|^2$, which weights each peak by its structure factor intensity. 

Fig.~\ref{Fig:ACOM_corr_match}f shows 2D slices of the 3D orientation plan, for the 5 diffraction patterns shown in Fig.~\ref{Fig:ACOM_corr_match}e. The in-plane rotational symmetry of each radial band is obvious for the low index zone axes, e.g. for the $[001]$ orientated crystal, the first row of the corresponding orientation plan consists of four spots which maintains the 4-fold symmetry of the diffraction pattern and can be indexed as $[020]$, $[200]$, $[0\overline{2}0]$ and $[\overline{2}00]$. The final step is to take the 1D Fourier transform along the $\phi_3$ axis in preparation for the Fourier correlation step defined in the next section.

%We repeat the orientation plan calculation for the number of phases that we are searching for in a given material.

% $(q_m,\phi_3,I_m)$.

\subsection*{Correlation Pattern Matching}

For each diffraction pattern measurement, we first measure the location and intensity of each Bragg disk by using the template matching procedure outlined by \cite{savitzky2021py4dstem}. The result is a set of $M$ experimental diffraction peaks defined by $(q_m,\gamma_m ,I_m)$. From these peaks, we calculate the sparse polar diffraction image $X(\phi_3, q_s)$ using the expression
\begin{eqnarray}
 \label{eq:image_polar}
    && X(\phi_3, q_s) = 
    \sum_{\{ q_m \,: \, |q_m - q_s| < \delta \}} 
    {q_m}^\gamma {I_m}^{\omega/2} \times
    \rm{max} \left\{
        1 - \right. 
    \\ % \nonumber
    &&
        \left. \frac{1}{\delta} \sqrt{
        (q_m - q_s)^2 + 
        \left[
        \rm{mod}(
        \phi_3 - \gamma_m + \pi, 2 \pi) - \pi
        \right]^2 {q_s}^2
        }, 0
    \right\}.
    \nonumber 
\end{eqnarray}
By default, we again use prefactors weighted by the peak radius and estimated peak amplitude given by the square root of the measured disk intensities.  However, if the dataset being analyzed contains a large number of different sample thicknesses, multiple scattering can cause strong oscillations in the peak amplitude values. As we will see in the simulations below, in these situations the best results may be achieved by setting $\omega=0$, i.e.\ ignoring peak intensity and weighting only by the peak radii. Note that in the diffraction image, the correlation kernel size $\delta$ again gives the azimuthal extent of the correlation signal. However, in Eq.~\ref{eq:image_polar} it also sets the range over which peaks are included in a given radial bin, and the fraction of the intensity assigned to each radial bin. 

Next, we calculate the correlation $C((\phi_1, \theta_2), \phi_3)$ of this image with the orientation plan using the expression
\begin{eqnarray}
    && C((\phi_1, \theta_2), \phi_3) 
    = 
    \sum_{q_s}
    \ift \left\{  \right.
    \nonumber \\
    &&
    \left.
        \ft\left\{ 
        {P((\phi_1, \theta_2), \phi_3, q_s)}
        \right\}^*
        \ft\left\{ 
        X(\phi_3, q_s)
        \right\}
    \right\},
    \nonumber 
\end{eqnarray}
where $\ft$ and $\ift$ are 1D forward and inverse fast Fourier transforms (FFTs) respectively along the $\phi_3$ direction, and the ${}^*$ operator represents taking the complex conjugate. We use this correlation over $\phi_3$ to efficiently calculate the in-plane rotation of the diffraction patterns. The maximum value in the correlogram will ideally correspond to the most probable orientation of the crystal. In order to account for mirror symmetry of the 2D diffraction patterns, we can also compute the correlation 
\begin{eqnarray}
    && C_{\rm{mirror}}((\phi_1, \theta_2), \phi_3) 
    = 
    \sum_{q_s}
    \ift \left\{  \right.
    \nonumber \\
    &&
    \left.
        \ft\left\{ 
        {P((\phi_1, \theta_2), \phi_3, q_s)}
        \right\}^*
        \ft\left\{ 
        X(\phi_3, q_s)
        \right\}^*
    \right\},
    \nonumber 
\end{eqnarray}
where the mirror operation is accomplished by taking the complex conjugate of $\ft\left\{X(\phi_3, q_s)\right\}$. For each zone axis $(\phi_1, \theta_2)$, we take the maximum value of $C$ and $C_{\rm{mirror}}$ in order to account for this symmetry. Figs.~\ref{Fig:ACOM_corr_match}g and h show 5 output correlograms, for the 5 diffraction patterns shown in Fig.~\ref{Fig:ACOM_corr_match}e. For each zone axis $(\phi_1, \theta_2)$, we have computed the maximum correlation value, which are plotted as a 2D array in Fig.~\ref{Fig:ACOM_corr_match}g. In each case, the highest value corresponds to the correct orientation. 

Fig.~\ref{Fig:ACOM_corr_match}h shows the correlation values along the $\phi_3$ axis, for the $(\phi_1, \theta_2)$ bins with the highest correlation value in Fig.~\ref{Fig:ACOM_corr_match}g. The symmetry of the correlation values in Fig.~\ref{Fig:ACOM_corr_match}h reflect the symmetry of the underlying patterns. For the $[0,0,1]$, $[0,1,1]$, and $[1,1,1]$, diffraction patterns, the in-plane angle $\phi_3$ correlation signals have 4-fold, 2-fold, and 6-fold rotational symmetry respectively. By contrast, the asymmetric diffraction patterns with zone axes $[1,1,3]$, and $[1,3,5]$ have only a single best in-plane orientation match.

\subsection*{Matching of Overlapping Diffraction Patterns}

In order to match multiple overlapping crystal signals, we have implemented an iterative detection process.  First, we use the above algorithm to determine the best fit orientation for a given pattern. Next, the forward diffraction pattern is calculated for this orientation. We then loop through all experimental peaks, and any within a user-specified deletion radius are removed from the pattern. Peaks which are outside of this radius, but within the correlation kernel size, have their intensities reduced by a factor defined by the linear distance between the experimental and simulated peaks divided by the distance between the correlation kernel size and the deletion radius. Then, the ACOM correlation matching procedure is repeated until the desired number of matches have been found, or no further orientations are found. Note that while we could update the correlation score after peak deletion, we output the original magnitude of the full pattern correlogram in order to accurately calculate the probability of multiple matches.

% \begin{equation}
%     C((\phi_1, \theta_2), \phi_3) 
%     = 
%     \ift \left\{
%         \ft\left\{ 
%         P((\phi_1, \theta_2), \phi_3, q_s)
%         \right\}
%         \ft\left\{ 
%         X(\phi_3, q_s)
%         \right\}
%     \right\}
% \end{equation}

% \subsection*{Simulations of Large Diffraction Libraries}

% \hl{ALEX R - ~1 paragraph summary of the diff library with needed citations}

\begin{figure}[htbp]
    \centering
    \includegraphics[width=3.2
    in]{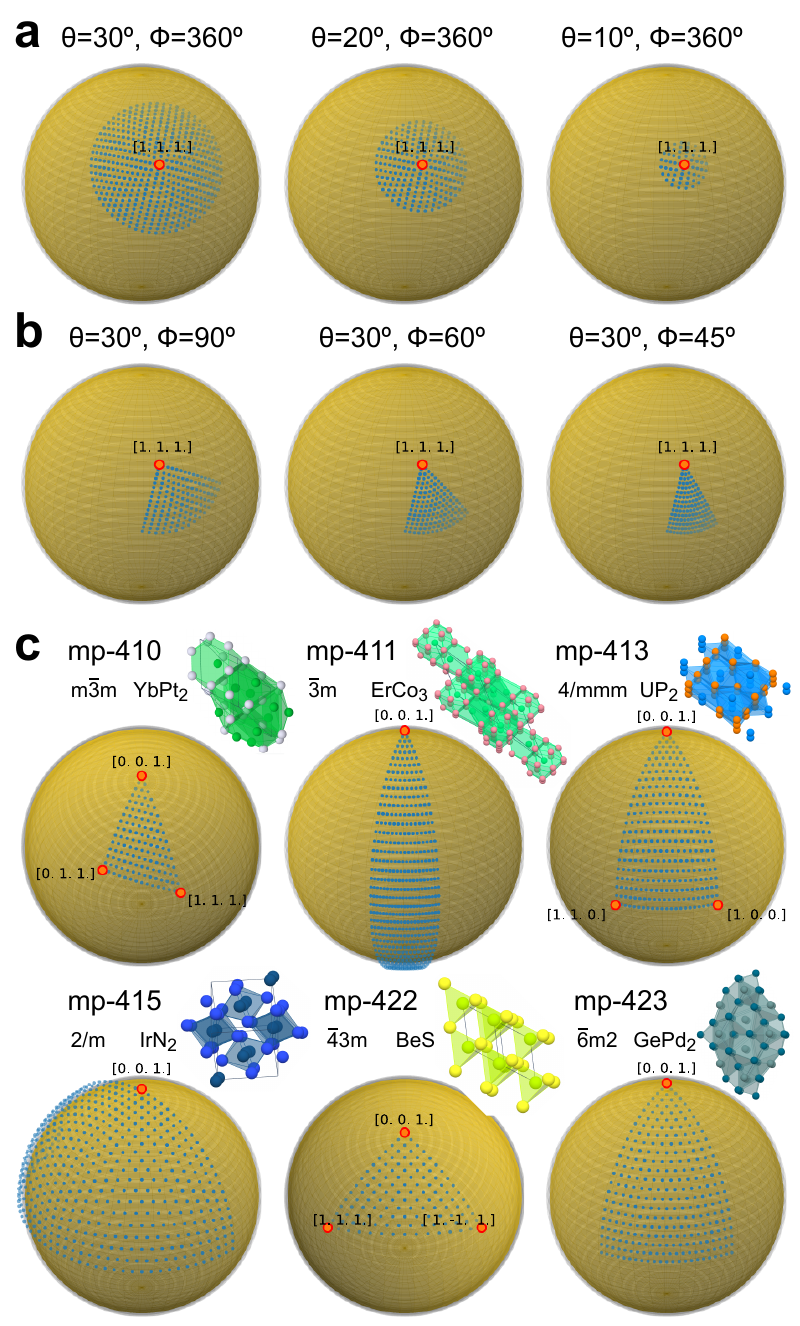}
    \caption{{\bf Examples of alternative orientation plan types in \pyFDSTEM{}}. Fiber texture examples where (a) orientations fully orbit around a single zone axis (the fiber axis), or (b) contain only a symmetry-reduced wedge of zone axes which orbit around a the fiber axis. (c) Examples of orientation plans generated directly from Materials Project entries \citep{jain2013commentary}, using pymatgen symmetries \citep{ong2013python}.}
    \label{Fig:ACOM_plans}
\end{figure}

\subsection*{ACOM Integration into \pyFDSTEM{}}

The ACOM pattern matching described has been implemented into the \pyFDSTEM{} python toolkit written by \cite{savitzky2021py4dstem}. A typical ACOM workflow starts with using \pyFDSTEM{} to import the 4D dataset and one or more images of the vacuum probe. We then use a correlation template matching procedure to find the positions of all diffracted disks at each probe position \citep{pekin2017optimizing}. We use the correlation intensity of each detected peak as an estimate of the peak's intensity. The resulting set of $M$ peaks defined by the values $(q_m, \gamma_m, I_m)$ are stored as a \emph{PointList} object in \pyFDSTEM{}. Because the number of peaks detected at each probe position can vary, we store the full set of all detected peaks in a \emph{PointListArray} object in \pyFDSTEM{}, which provides an interface to the ragged structured numpy data.

Most experimental datasets contain some degree of ellipticity, and the absolute pixel size must be calibrated. We perform these corrections on the set of measured diffraction disks using the \pyFDSTEM{} calibration routines defined by \cite{savitzky2021py4dstem}. We know that the correlation approach is relatively robust against both ellipticity and small errors in the reciprocal space pixel size. However, precise phase mapping may require us to distinguish between crystals with similar lattice parameters; these experiments will require accurate calibration.

We perform ACOM in \pyFDSTEM{} by first creating a \emph{Crystal} object, either by specifying the atomic basis directly, or by using the pymatgen package \citep{ong2013python} to import structural data from crystallographic  information files (CIF), or the Materials Project database \citep{jain2013commentary}. The \emph{Crystal} object is used to calculate the structure factors, and generate an orientation plan. The final step is to use the orientation plan to determine the best match (or matches) for each probe position, from the list of calibrated diffraction peaks.  If the sample contains multiple phases, we perform the orientation plan calculation and correlation matching for each unique crystal structure.

%for the number of phases that we are searching for in a given material.

In addition to specifying the orientation plan spanning 3 vectors as in Fig.~\ref{Fig:ACOM_corr_match}, we define additional methods to describe the space of possible orientations. One such example is \emph{fiber texture}, where we assume the crystals are all orientated near a single zone axis known as the fiber axis, shown in Figs.~\ref{Fig:ACOM_plans}a and b. We can vary the angular range of zone axes included away from the fiber axis as in Fig.~\ref{Fig:ACOM_plans}a, as well as choose the azimuthal range around this axis as in Fig.~\ref{Fig:ACOM_plans}b to account for symmetry around the fiber axis. Alternatively, an ``automatic'' option is provided, which uses pymatgen to determine the symmetry of the structure and automatically choose the span of symmetrically unique zone axes which should be included in the orientation plan, based on the point group symmetry \cite{de2003introduction}. This is shown for a selection of different Materials Project database entries in Fig.~\ref{Fig:ACOM_plans}c.

\subsection*{Simulations of Diffraction Patterns from Thick Samples}

%  To test the performance of our methods on diffraction patterns // which contain significant multiple scattering, we have 

One important metric for the performance of an orientation mapping algorithm is how well it performs when the diffraction patterns contain significant amounts of multiple scattering. We have therefore used our ACOM algorithm to measure the orientation of simulated diffraction patterns from samples tilted along many directions, over a wide range of thicknesses. We performed these simulations using the multislice algorithm \citep{cowley1957scattering}, and methods defined by \cite{kirkland2020advanced} and \cite{ophus2017fast}. These methods are implemented in the \prismatic{} simulation code by \cite{dacosta2021prismatic}. The diffraction patterns were generated using a acceleration potential of 300 keV, a 0.5 mrad convergence angle, with real space and reciprocal pixel sizes of 0.05 $\rm{\AA}$ and 0.01 $\rm{\AA}^{-1}$ respectively, with 4 frozen phonons. In total we have simulated 3750 diffraction patterns from Cu, Ag, and Au fcc crystals, over 25 zone axes ($[0,0,1]$ to $[3,4,4]$ excluding symmetrically redundant reflections) and thicknesses up to 100 nm with a 2 nm step size. These diffraction patterns were generated using the simulation pipeline and database defined by \cite{rakoski2021database}.
%\hl{add a reference to the git repo in lieu SI?}. 

\subsection*{Chemical Synthesis of Twisted AuAgPd Nanowires}

The performed synthesis was modified from a known method given by \cite{wang2011}. We prepared the following solutions: \SI{500}{\milli\Molar} PVP (MW 40,000) in DMF, \SI{50}{\milli\Molar} HAuCl$_{4}$ in DMF, \SI{50}{\milli\Molar} AgNO$_{3}$ in MilliQ water, and \SI{400}{\milli\Molar} L-ascorbic acid in MilliQ water. We created the reaction solution in a 4 mL vial (washed 3x with MilliQ water and acetone) by mixing \SI{800}{\micro\liter} DMF, \SI{100}{\micro\liter} PVP, \SI{20}{\micro\liter} HAuCl$_{4}$, and \SI{20}{\micro\liter} AgNO$_{3}$. We vortexed the solution, then added \SI{100}{\micro\liter} of L-ascorbic acid solution drop-wise to the mixture while gently swirling. At this point, the color changed from pale yellow to clear. We left the solution at room temperature for 7 days, at which point the solution was light brown/purple. The primary product of this reaction was straight, ultrathin Au-Ag nanowires (2 nm in diameter).

To twist the underlying ultrathin Au-Ag nanowires, we prepared solutions of \SI{1.875}{\milli\Molar} L-ascorbic acid and \SI{2}{\milli\Molar} H$_{2}$PdCl$_{2}$ in MilliQ water. In a 4 mL vial (3x washed with MilliQ water/acetone), we added \SI{50}{\micro\liter} of the Au-Ag reacted solution to \SI{640}{\micro\liter} of the L-ascorbic acid solution. Finally, we added \SI{60}{\micro\liter} of the H$_{2}$PdCl$_{4}$ solution and allowed the sample to incubate for at least 30 minutes. We purified the reaction solution by centrifuging the product down at 7500 rpm for 4 minutes. We decanted the supernatant, and then rinsed the reaction with MilliQ water 3 times and re-dispersed in MilliQ water. We prepared TEM samples of this material by depositing \SI{10}{\micro\liter} of purified nanowire solution onto 400 mesh formvar/ultrathin carbon grids.

\subsection*{4D-STEM Experiments with Patterned Apertures}

We collected the experimental data using a double aberration-corrected modified FEI Titan 80-300 microscope (the TEAM I instrument at the National Center for Electron Microscopy within Lawrence Berkeley  National Laboratory). This microscope is equipped with a Gatan K3 detector and Continuum spectrometer, and was set to collect diffraction patterns integrated over 0.05 seconds, with 4x binning. We used an accelerating voltage of 300 keV, an energy slit of 20 eV, and a spot size of 6. We used a \SI{10}{\micro\meter} bullseye aperture to form the STEM probe in order to improve detection precision of the Bragg disks \citep{zeltmann2020}. We used a convergence semiangle of \SI{2}{\milli\radian}, with a camera length of 1.05 m. We recorded the experimental dataset using a step size of \SI{5}{\angstrom}, with a total of 286 and 124 steps in the x and y directions.

\begin{figure}[htbp]
    \centering
    \includegraphics[width=3.2 in]{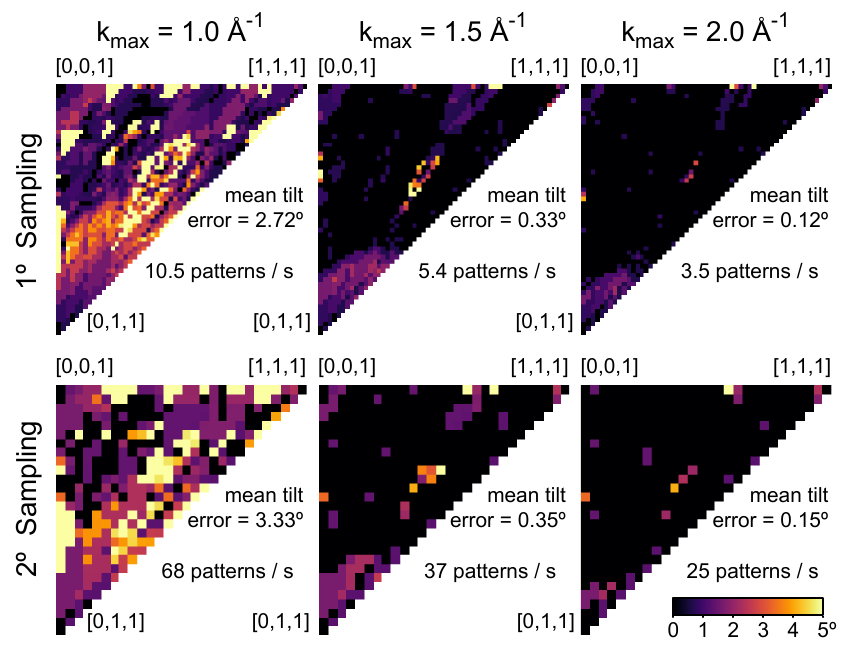}
    \caption{{\bf Zone axis misorientation as a function of sampling and maximum scattering angle for kinematical simulations.} The mean tilt error and number of patterns matched per second are shown inset for each panel.}
    \label{Fig:ACOM_kinematical}
\end{figure}

\begin{figure*}[htbp]
    \centering
    \includegraphics[width=6.0 in]{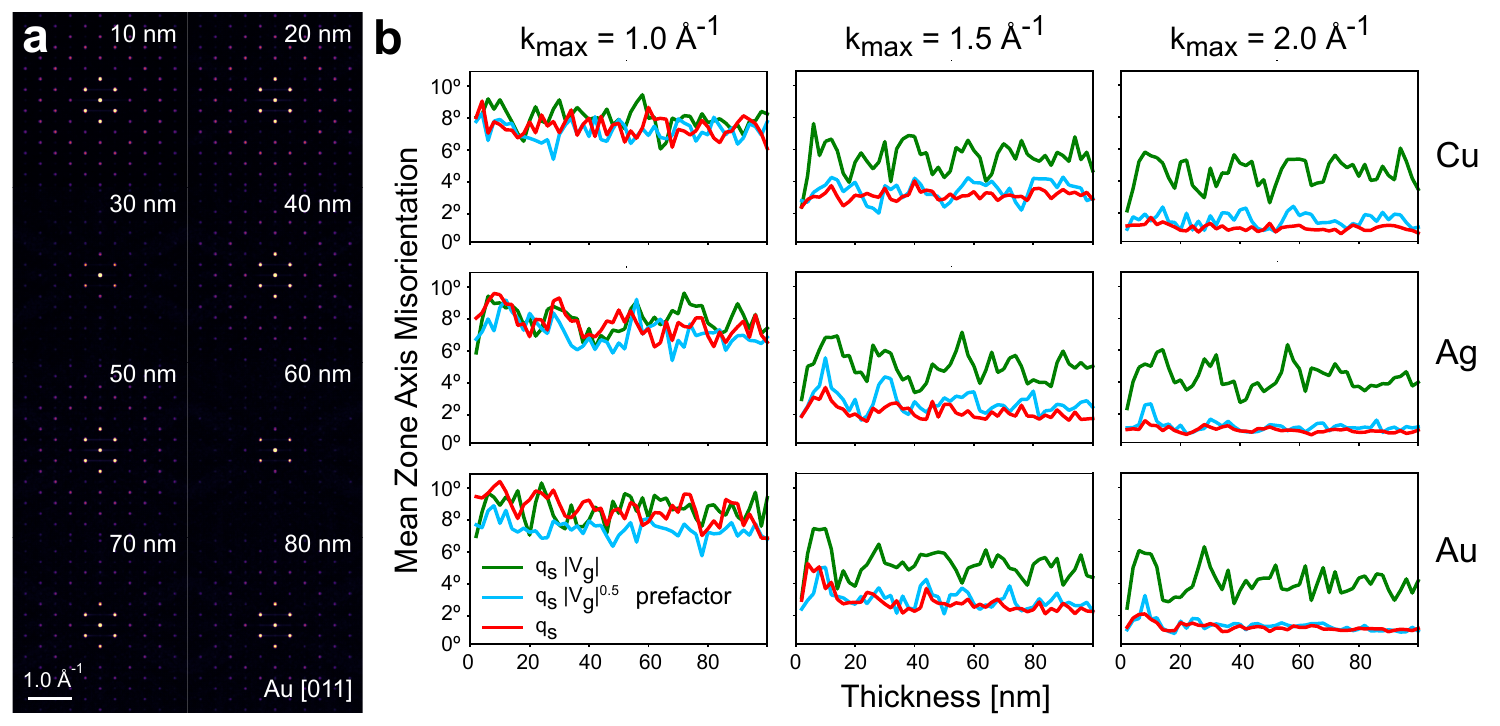}
    \caption{{\bf Dynamical simulated diffraction patterns.} (a) Example diffraction patterns for Au oriented to the [011] zone axis for 10-80 nm thick slices. (b) Plots showing the mean zone axis misorientation in degrees as a function of thickness for Cu, Ag, and Au. Each plot shows the errors for correlation prefactors of $q_{\rm{s}} |V_{\rm{g}}|$ (red) and $q_{\rm{s}}$ (blue).}
    \label{Fig:dynamical_diffractions}
\end{figure*}

\section*{Results and Discussion}

\subsection*{ACOM of Kinematical Calculated Diffraction Patterns}

For the first test of our correlation method, we applied it to the same patterns calculated to generate an orientation plan for fcc Au. Next, we measured the calculation time and angular error between the measured and ground truth zone axes for each pattern. These results are plotted in Fig.~\ref{Fig:ACOM_kinematical} for 3 different maximum scattering angles $k_{\rm{max}}$, and angular sampling of $1^\circ$ and $2^\circ$.

The results in Fig.~\ref{Fig:ACOM_kinematical} show that the angular error in zone axis orientation is relatively insensitive to the angular sampling. However, the angular error drops by a factor of 10 from approximately $\approx 3^\circ$ to $\approx 0.3^\circ$ when increasing the maximum scattering angle included from $k_{\rm{max}} = 1 \, {\rm{\AA}}^{-1}$ to $1.5 \, {\rm{\AA}}^{-1}$, and by another factor of 2-3 when increasing $k_{\rm{max}}$ to $2 \, {\rm{\AA}}^{-1}$. This is unsurprising, as examining Fig.~\ref{Fig:ACOM_corr_match}e shows that there is a large increase in the number of visible Bragg spots outside of $k_{\rm{max}} = 1 \, {\rm{\AA}}^{-1}$, and because Bragg disks at higher scattering angles provide better angular precision relative to low $k$ disks. This result emphasizes the importance of collecting as wide of angular range as possible when performing orientation matching of 4D-STEM data.

The inset calculation times reported are for the single-threaded ACOM implementation in \pyFDSTEM, running in Anaconda \citep{anaconda} on a laptop with an Intel Core i7-10875H processor, running at 2.30 GHz. The calculation times can be increased by an order of magnitude or more when running in parallel, or by using a GPU to perform the matrix multiplication and Fourier transform steps.

\subsection*{ACOM of Dynamical Simulated Diffraction Patterns}

% (top, middle and bottom rows respectively). Left, middle and right columns correspond to $k_{\{rm{max}}$ values of 1.0, 1.5, and 2.0.

In diffraction experiments using thick specimens, the electron beam can scatter multiple times, a phenomenon known as dynamical diffraction. This effect is especially pronounced in diffraction experiments along low index zone axes, where the diffracted peak intensities oscillate a function of thickness. In order to test the effect of oscillating peak intensities on our ACOM method, we have simulated diffraction patterns for Cu, Ag, and Au fcc crystals, along multiple zone axes. Some example diffraction patterns for the [011] zone axis of Au are plotted in Fig.~\ref{Fig:dynamical_diffractions}a. We see that all diffraction spots have intensities which oscillate multiple times as a function of thickness. 

We performed ACOM by generating orientation plans with an angular sampling of $2^\circ$, a correlation kernel size of 0.08 $\rm{\AA}^{-1}$, and maximum scattering angles of $k_{\rm{max}}$ = 1.0, 1.5, and 2.0 $\rm{\AA}^{-1}$. We kept the radial prefactor of weighting set to $\gamma $ = 1, and tested peak amplitude prefactors of $\omega$ = 1.0, 0.5 and 0.0. The average zone axis angular misorientation as a function of thickness is plotted in Fig.~\ref{Fig:dynamical_diffractions}b. In total, we performed orientation matching on 3750 diffraction patterns, and a total of 33750 correlation matches on a  workstation with an AMD Ryzen Threadripper 3960X CPU (2.2 GHz, baseclock). The typical number of patterns matched per second were of ~80-90, 45-55 and 25-30 patterns/s for $k_{\rm{max}}$ values of 1.0, 1.5, and 2.0 $\rm{\AA}^{-1}$ respectively. 

As expected, the errors are higher than those achieved under kinematic conditions, and the trend for smaller errors with larger $k_{\rm{max}}$ is also preserved (mean errors of 7.25$^\circ$, 3.09$^\circ$ and 1.39$^\circ$ for $k_{\rm{max}}$ values 1.0, 1.5 and 2.0 $\rm{\AA}^{-1}$ respectively, $\gamma $ = 1, $\omega$ = 0.25). We did not observe any dependence of the orientation accuracy on the simulation thickness. Despite the correlation prefactor $|Vg|$ performing well for the examples shown in Fig.~\ref{Fig:ACOM_corr_match}, for the dynamical diffraction simulations along zone axes it was out-performed by prefactors of both $\sqrt{|V_g|}$ ($\omega$ = 0.5) and omitting the peak amplitude prefactor altogether ($\omega$ = 0). We therefore suggest that when mapping samples with a large range of thicknesses, or many crystals aligned to low index zone axes, the position of the diffracted peaks is significantly more important than their amplitudes or intensities. One possible method to increase the accuracy while using higher amplitude prefactors is to perform an experiment which recovers more kinematical values for the diffracted peak intensities, for example by precessing the electron beam when recording diffraction patterns \citep{midgley2015precession, jeong2021automated}. We note that there is likely no global optimal choice of orientation mapping hyperparameters for all materials and thicknesses, and this may be a worthwhile topic for future investigations.

\subsection*{4D-STEM ACOM of Twisted AuAgPd Nanowires}

We have tested our ACOM algorithm with a 4D-STEM dataset collected for an AuAgPd nanowires. An image of the vacuum bullseye STEM probe is shown in Fig.~\ref{Fig:exp_structure}a. For each detector pixel, we have calculated the maximum value across all STEM probe positions to generate a \emph{maximum diffraction pattern}, shown in Fig.~\ref{Fig:exp_structure}b. The beamstop used to block the center beam is visible, as well as various crystalline diffraction rings out to approximately 1.4 ${\rm{\AA}}^{-1}$.

\begin{figure*}[htbp]
    \centering
    \includegraphics[width=6.4 in]{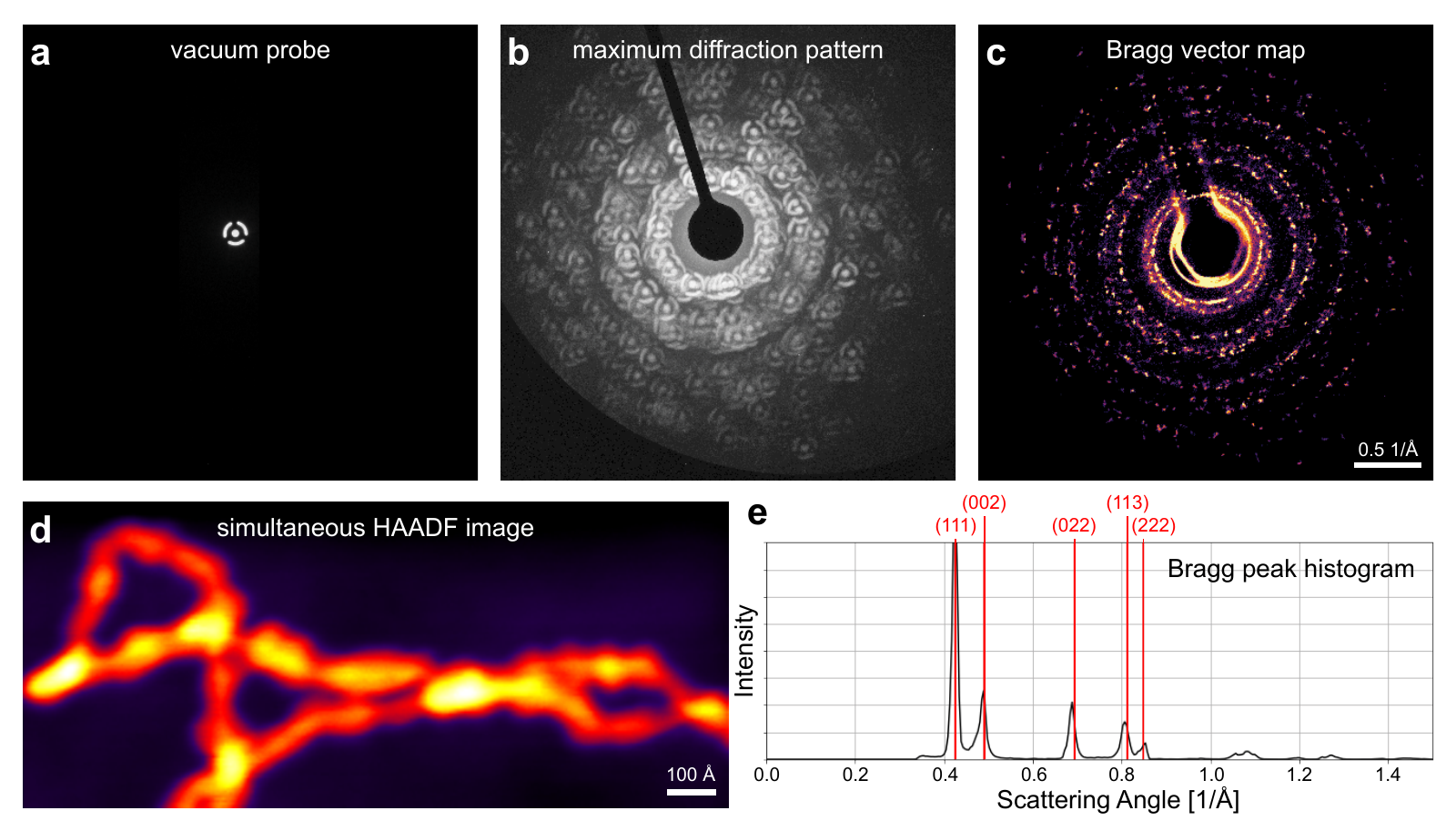}
    \caption{{\bf 4D-STEM scan of twisted polycrystalline AuAgPd nanowires.} (a) Diffraction image of probe over vacuum, showing bullseye pattern. (b) Maximum of each pixel in diffraction space over all probe positions. (c) Histogram of all peak locations detected by correlation in \pyFDSTEM{} of (a) with each pattern included in (b). (d) HAADF-STEM image of the sample. (e) 1D histogram of scattering vectors, with fcc AuAg inverse plane spacings overlaid.}
    \label{Fig:exp_structure}
\end{figure*}

After performing the correlation peak finding algorithm in \pyFDSTEM, we have an estimated position and intensity of all detected Bragg peaks. A 2D histogram of these peaks, known as a Bragg vector map, is plotted in Fig.~\ref{Fig:exp_structure}c. Sharp polycrystalline diffraction rings are clearly visible, as well as false positives generated by the beamstop edge. These false positives were manually removed by using a mask generated from an image of the beamstop. A high angle annular dark field (HAADF) image was simultaneously recorded during the 4D-STEM data collection, which is shown in Fig.~\ref{Fig:exp_structure}d.

The final experimental pre-processing steps are to calibrate the diffraction pattern center, the elliptical distortions, and the absolute pixel size. We performed these steps by fitting an ellipse to the $(022)$ diffraction ring, and by assuming a lattice constant of $4.08 \, \rm{\AA}$, corresponding to the fcc Au structure \citep{maeland1964lattice}. This process is explained in more detail by \cite{savitzky2021py4dstem}. We assumed that the Ag lattice constant is similar to that of Au. Despite the presence of Pd in the nanowires, there was no significant presence of secondary grains corresponding to the smaller lattice of fcc Pd grains. An intensity histogram of the corrected Bragg peak scattering angles are shown in Fig.~\ref{Fig:exp_structure}e. We have overlaid the 5 smallest scattering angles of Au on Fig.~\ref{Fig:exp_structure}e to show the accuracy of the correction.

\begin{figure*}[htbp]
    \centering
    \includegraphics[width=6.4in]{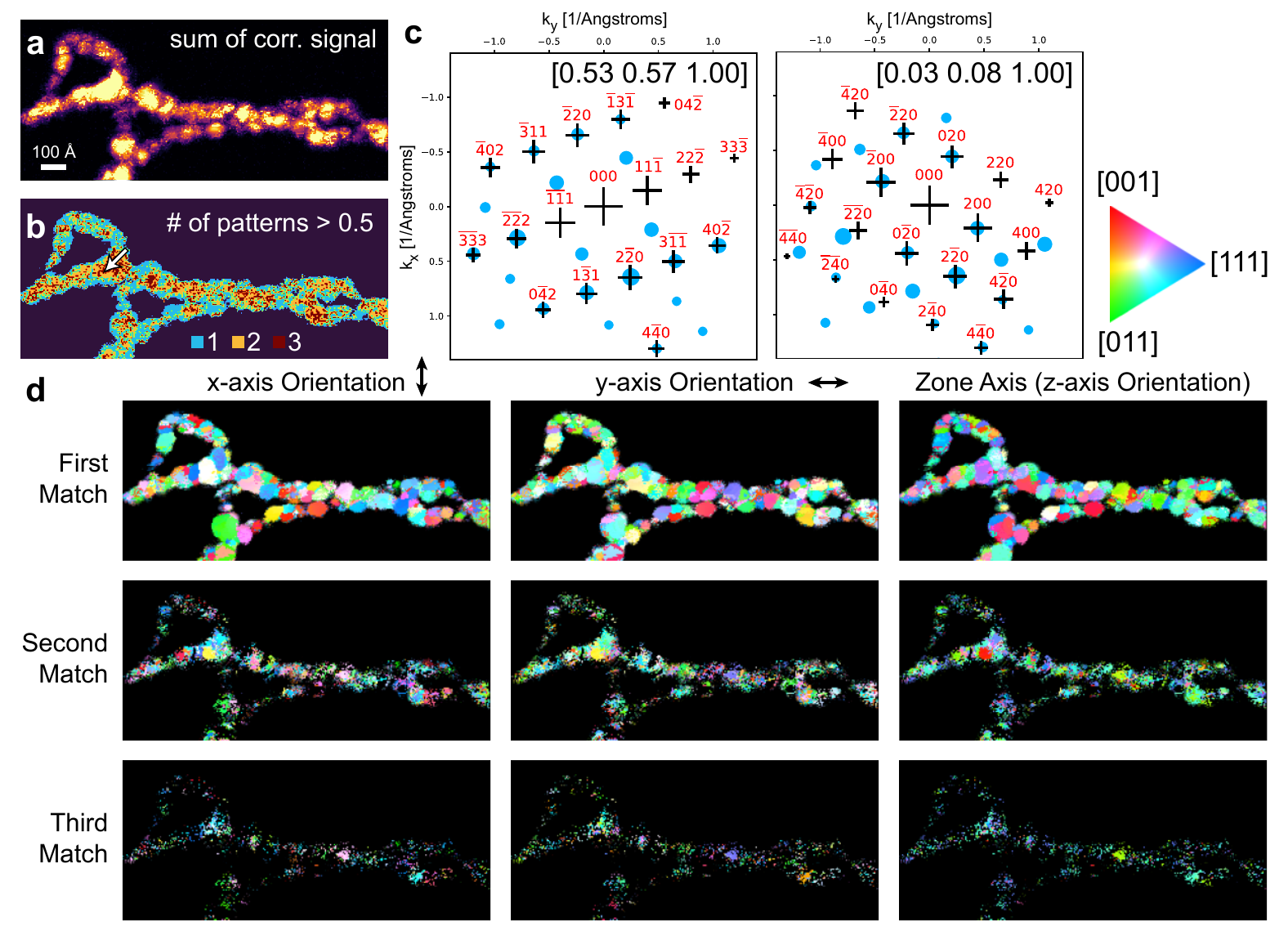}
    \caption{{\bf Orientation mapping of polycrystalline AuAgPd nanowires.} (a) Total of measured correlation signal for each probe position. (b) Estimated number of patterns indexed for each probe position. (c) Example of 2 orientations indexed from a single diffraction pattern, collected at the position indicated by the arrow shown in (b). (d) Orientation maps of the 3 highest correlation signals for each probe position. Legend shown above.}
    \label{Fig:ACOM_AuAgPd}
\end{figure*}

We have performed ACOM on the AuAgPd nanowire sample, with the results shown in Fig.~\ref{Fig:ACOM_AuAgPd} shown for up to 3 matches for each diffraction pattern. For each probe position, the sum of the maximum detected correlation signals for up to three matches are shown in Fig.~\ref{Fig:ACOM_AuAgPd}a. The structure is in good agreement with Fig.~\ref{Fig:exp_structure}, though with additional modulations due to some grains generating more diffraction signal than others. Using a correlation intensity threshold of 0.5, we have plotted the number of matching patterns in Fig.~\ref{Fig:ACOM_AuAgPd}b. The threshold of 0.5 was arbitrary chosen as a lower bound for a potential match. Examples of 2 matches to a single diffraction pattern are plotted in Fig.~\ref{Fig:ACOM_AuAgPd}c. In this figure the correlation score for the first matched pattern was higher than the second. The second match found shows some deformation between the measured and simulated Bragg peak positions, and matches less peaks. It therefore produces a lower correlation score, which can be used to threshold the results as in Fig.~\ref{Fig:ACOM_AuAgPd}d.

% \hl{Additionally, the second match shows some deformation of the lattice vectors from the ideal structure, probably indicating that there is a crystallographic relationship between these patterns. } \hl{SEZ-What does `deformation of the lattice vectors' mean here, and how can you tell that form the match intensities??}

Fig.~\ref{Fig:ACOM_AuAgPd}d shows the 3D orientations for all probe positions, with the 3 best matches shown. Each image is masked by the total correlation signal, so that low correlation values are colored black. Almost every diffraction pattern with Bragg disks detected was indexed for at least one orientation with high confidence. Additionally, the patterns are very consistent, with a large number of adjacent probe positions recording the same orientation. Some secondary grains are also clearly visible in the second-best match, while very few patterns have been assigned a third match with high confidence. 

\begin{figure}[htbp]
    \centering
    \includegraphics[width=3.4in]{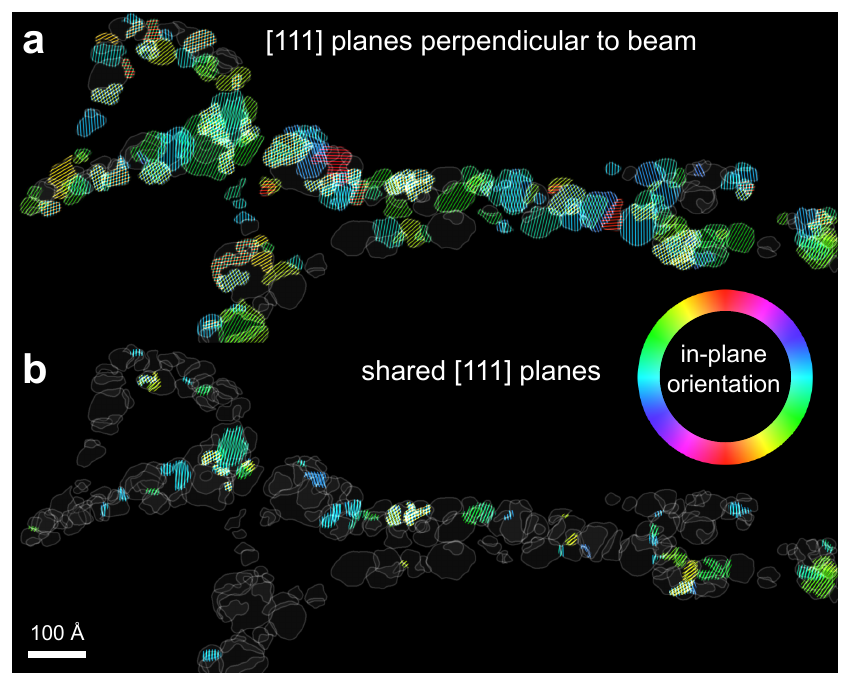}
    \caption{{\bf Orientation analysis of grains in AuAgPd nanowires.} (a) Crystal grains, with in-plane (111) planes colored by orientation. (b) (111) planes shared by two overlapping grains.}
    \label{Fig:ACOM_111_twins}
\end{figure}

In order to investigate the grain organization of the AuAgPd nanowires, we have performed clustering analysis on the orientation maps. Grains with similar orientations have been clustered together by looping through each probe position and comparing its orientation to its neighbors. Grains with at least 10 contiguous probe positions are shown in Fig.~\ref{Fig:ACOM_111_twins}a. $(111)$ planes which lie in the image plane are overlaid onto the grain strucure, colored by their orientation. Confirming our observations in Fig.~\ref{Fig:ACOM_AuAgPd}d, only a few grains with substantial overlap were reliably identified. This might be due to the low thickness of the sample (only a single grain along the beam direction), some grains not being oriented close enough to a zone axis to be detected, or multiple scattering deviations in the diffracted signal. There is a noticable bias in the orientation of the $(111)$ planes, which tend to be oriented horizontally near the growth direction of the nanowires.

%There does not appear to be any noticeable pattern in the overall orientations of the grains.

One hypothesis for the growth mode of these twisted nanowires is that adjacent grains are connected by $(111)$ twin planes, forming local helical structures to give the observed twisted structures. To test this hypothesis, we determined the position of $(111)$ planes from Fig.~\ref{Fig:ACOM_111_twins}a which are shared by two overlapping grains. Fig.~\ref{Fig:ACOM_111_twins}b shows the location of these shared $(111)$ planes (with plane normal differences below $8^\circ$), colored by the normal vector of the plane. Many shared $(111)$ planes were detected, most with normal vectors aligned to the wire growth direction. These observations support the hypothesis that these nanowires are composed of grains connected by $(111)$ twin planes.

%This is reasonable since in order to detect these shared planes, we require the grains on either side to be oriented along a low enough index zone axis. If both grains have a $(111)$ plane normal in the sample plane, the odds of both being in a good diffraction condition are increased. 

These experimental observations demonstrate the efficacy of our ACOM method. In order to improve these results, we will need to collect diffraction data with a wider angular range. This can be achieved by using precession electron diffraction \citep{rouviere2013improved}, multibeam electron diffraction \cite{hong2021multibeam}, or by tilting the sample or beam and recording multiple 4D-STEM datasets \citep{meng2016three}.

\section*{Conclusion}

We have introduced an efficient and accurate method to perform automated crystal orientation mapping, using a sparse correlation matching procedure. We have implemented our methods into the open source \pyFDSTEM{} toolkit, and demonstrated the accuracy of our method using simulated diffraction patterns. We also applied ACOM to an experimental scan of a complex helical polycrystalline nanowire, where we were able to identify shared twin planes between adjacent grains which may be responsible for the twisted helical geometry. All of our methods have been made freely available to the microscopy community as open source codes. We believe that our implementation of ACOM is efficient and accurate enough to be Incorporated into automated online TEM software \citep{spurgeon2021towards}. In the future, we will improve our ACOM method using machine learning methods \cite{munshi2021ml}, and we will extend our ACOM methods to include multibeam electron diffraction experiments \citep{hong2021multibeam}.

% and we will extend these methods in future parts of this manuscript. 

\section*{Source Code and Data Availability}

All code used in this manuscript is available on the \href{https://github.com/py4dstem/py4DSTEM/tree/acom}{py4DSTEM GitHub repository}, and the tutorial notebooks are available on the \href{https://github.com/py4dstem/py4DSTEM_tutorials/tree/main/notebooks/acom}{py4DSTEM tutorial repository}. All simulated and experimental 4D-STEM datasets are available at [links will be added after publication].

%  (currently on the ACOM branch, will be moved to main py4DSTEM repo after publication)

\section*{Acknowledgements}

We thank Karen Bustillo for helpful discussions. CO acknowledges support of a US Department of Energy Early Career Research Award. SEZ was supported by the National Science Foundation under STROBE Grant no. DMR 1548924. AB, BHS, and \pyFDSTEM{} development are supported by the Toyota Research Institute. AR is supported by the 4D Data Distillery project, funded by the US Department of Energy. Work at the Molecular Foundry was supported by the Office of Science, Office of Basic Energy Sciences, of the U.S. Department of Energy under Contract No. DE-AC02-05CH11231. This research used resources of the National Energy Research Scientific Computing Center (NERSC), a U.S. Department of Energy Office of Science User Facility located at Lawrence Berkeley National Laboratory, operated under Contract No. DE-AC02-05CH11231.

\section*{References}
\bibliography{refs}    

%aipnauth4-2.bst 2018-12-27 (MD) hand-edited version of apsauth4-1.bst
%Control: author (9) reversed initials
\begin{thebibliography}{66}%
\makeatletter
\providecommand \@ifxundefined [1]{%
 \@ifx{#1\undefined}
}%
\providecommand \@ifnum [1]{%
 \ifnum #1\expandafter \@firstoftwo
 \else \expandafter \@secondoftwo
 \fi
}%
\providecommand \@ifx [1]{%
 \ifx #1\expandafter \@firstoftwo
 \else \expandafter \@secondoftwo
 \fi
}%
\providecommand \natexlab [1]{#1}%
\providecommand \enquote  [1]{``#1''}%
\providecommand \bibnamefont  [1]{#1}%
\providecommand \bibfnamefont [1]{#1}%
\providecommand \citenamefont [1]{#1}%
\providecommand \href@noop [0]{\@secondoftwo}%
\providecommand \href [0]{\begingroup \@sanitize@url \@href}%
\providecommand \@href[1]{\@@startlink{#1}\@@href}%
\providecommand \@@href[1]{\endgroup#1\@@endlink}%
\providecommand \@sanitize@url [0]{\catcode `\\12\catcode `\$12\catcode
  `\&12\catcode `\#12\catcode `\^12\catcode `\_12\catcode `\%12\relax}%
\providecommand \@@startlink[1]{}%
\providecommand \@@endlink[0]{}%
\providecommand \url  [0]{\begingroup\@sanitize@url \@url }%
\providecommand \@url [1]{\endgroup\@href {#1}{\urlprefix }}%
\providecommand \urlprefix  [0]{URL }%
\providecommand \Eprint [0]{\href }%
\providecommand \doibase [0]{https://doi.org/}%
\providecommand \selectlanguage [0]{\@gobble}%
\providecommand \bibinfo  [0]{\@secondoftwo}%
\providecommand \bibfield  [0]{\@secondoftwo}%
\providecommand \translation [1]{[#1]}%
\providecommand \BibitemOpen [0]{}%
\providecommand \bibitemStop [0]{}%
\providecommand \bibitemNoStop [0]{.\EOS\space}%
\providecommand \EOS [0]{\spacefactor3000\relax}%
\providecommand \BibitemShut  [1]{\csname bibitem#1\endcsname}%
\let\auto@bib@innerbib\@empty
%</preamble>
\bibitem [{ana(2020)}]{anaconda}%
  \BibitemOpen
  \href {https://docs.anaconda.com/} {\enquote {\bibinfo {title} {Anaconda
  software distribution},}\ } (\bibinfo {year} {2020})\BibitemShut {NoStop}%
\bibitem [{\citenamefont {Borchardt-Ott}(2011)}]{borchardt2011crystallography}%
  \BibitemOpen
  \bibfield  {author} {\bibinfo {author} {\bibnamefont {Borchardt-Ott},
  \bibfnamefont {W.}},\ }\href@noop {} {\emph {\bibinfo {title}
  {Crystallography: an introduction}}}\ (\bibinfo  {publisher} {Springer
  Science \& Business Media},\ \bibinfo {year} {2011})\BibitemShut {NoStop}%
\bibitem [{\citenamefont {Brunetti}\ \emph {et~al.}(2011)\citenamefont
  {Brunetti}, \citenamefont {Robert}, \citenamefont {Bayle-Guillemaud},
  \citenamefont {Rouviere}, \citenamefont {Rauch}, \citenamefont {Martin},
  \citenamefont {Colin}, \citenamefont {Bertin},\ and\ \citenamefont
  {Cayron}}]{brunetti2011confirmation}%
  \BibitemOpen
  \bibfield  {author} {\bibinfo {author} {\bibnamefont {Brunetti},
  \bibfnamefont {G.}}, \bibinfo {author} {\bibnamefont {Robert}, \bibfnamefont
  {D.}}, \bibinfo {author} {\bibnamefont {Bayle-Guillemaud}, \bibfnamefont
  {P.}}, \bibinfo {author} {\bibnamefont {Rouviere}, \bibfnamefont {J.}},
  \bibinfo {author} {\bibnamefont {Rauch}, \bibfnamefont {E.}}, \bibinfo
  {author} {\bibnamefont {Martin}, \bibfnamefont {J.}}, \bibinfo {author}
  {\bibnamefont {Colin}, \bibfnamefont {J.}}, \bibinfo {author} {\bibnamefont
  {Bertin}, \bibfnamefont {F.}}, and\ \bibinfo {author} {\bibnamefont {Cayron},
  \bibfnamefont {C.}},\ }\bibfield  {title} {\enquote {\bibinfo {title}
  {Confirmation of the domino-cascade model by {LiFePO$_4$/FePO$_4$} precession
  electron diffraction},}\ }\href@noop {} {\bibfield  {journal} {\bibinfo
  {journal} {Chemistry of Materials}\ }\textbf {\bibinfo {volume} {23}},\
  \bibinfo {pages} {4515--4524} (\bibinfo {year} {2011})}\BibitemShut {NoStop}%
\bibitem [{\citenamefont {Bustillo}\ \emph {et~al.}(2021)\citenamefont
  {Bustillo}, \citenamefont {Zeltmann}, \citenamefont {Chen}, \citenamefont
  {Donohue}, \citenamefont {Ciston}, \citenamefont {Ophus},\ and\ \citenamefont
  {Minor}}]{bustillo20214d}%
  \BibitemOpen
  \bibfield  {author} {\bibinfo {author} {\bibnamefont {Bustillo},
  \bibfnamefont {K.~C.}}, \bibinfo {author} {\bibnamefont {Zeltmann},
  \bibfnamefont {S.~E.}}, \bibinfo {author} {\bibnamefont {Chen}, \bibfnamefont
  {M.}}, \bibinfo {author} {\bibnamefont {Donohue}, \bibfnamefont {J.}},
  \bibinfo {author} {\bibnamefont {Ciston}, \bibfnamefont {J.}}, \bibinfo
  {author} {\bibnamefont {Ophus}, \bibfnamefont {C.}}, and\ \bibinfo {author}
  {\bibnamefont {Minor}, \bibfnamefont {A.~M.}},\ }\bibfield  {title} {\enquote
  {\bibinfo {title} {{4D-STEM} of beam-sensitive materials},}\ }\href@noop {}
  {\bibfield  {journal} {\bibinfo  {journal} {Accounts of Chemical Research}\
  ,\ \bibinfo {pages} {860--866}} (\bibinfo {year} {2021})}\BibitemShut
  {NoStop}%
\bibitem [{\citenamefont {Castro-M{\'e}ndez}, \citenamefont {Hidalgo},\ and\
  \citenamefont {Correa-Baena}(2019)}]{castro2019role}%
  \BibitemOpen
  \bibfield  {author} {\bibinfo {author} {\bibnamefont {Castro-M{\'e}ndez},
  \bibfnamefont {A.-F.}}, \bibinfo {author} {\bibnamefont {Hidalgo},
  \bibfnamefont {J.}}, and\ \bibinfo {author} {\bibnamefont {Correa-Baena},
  \bibfnamefont {J.-P.}},\ }\bibfield  {title} {\enquote {\bibinfo {title} {The
  role of grain boundaries in perovskite solar cells},}\ }\href@noop {}
  {\bibfield  {journal} {\bibinfo  {journal} {Advanced Energy Materials}\
  }\textbf {\bibinfo {volume} {9}},\ \bibinfo {pages} {1901489} (\bibinfo
  {year} {2019})}\BibitemShut {NoStop}%
\bibitem [{\citenamefont {Cowley}\ and\ \citenamefont
  {Moodie}(1957)}]{cowley1957scattering}%
  \BibitemOpen
  \bibfield  {author} {\bibinfo {author} {\bibnamefont {Cowley}, \bibfnamefont
  {J.~M.}}and\ \bibinfo {author} {\bibnamefont {Moodie}, \bibfnamefont
  {A.~F.}},\ }\bibfield  {title} {\enquote {\bibinfo {title} {The scattering of
  electrons by atoms and crystals. {I}. a new theoretical approach},}\
  }\href@noop {} {\bibfield  {journal} {\bibinfo  {journal} {Acta
  Crystallographica}\ }\textbf {\bibinfo {volume} {10}},\ \bibinfo {pages}
  {609--619} (\bibinfo {year} {1957})}\BibitemShut {NoStop}%
\bibitem [{\citenamefont {De~Graef}(2003)}]{de2003introduction}%
  \BibitemOpen
  \bibfield  {author} {\bibinfo {author} {\bibnamefont {De~Graef},
  \bibfnamefont {M.}},\ }\href@noop {} {\emph {\bibinfo {title} {Introduction
  to conventional transmission electron microscopy}}}\ (\bibinfo  {publisher}
  {Cambridge university press},\ \bibinfo {year} {2003})\BibitemShut {NoStop}%
\bibitem [{\citenamefont {Dederichs}\ \emph {et~al.}(1978)\citenamefont
  {Dederichs}, \citenamefont {Lehmann}, \citenamefont {Schober}, \citenamefont
  {Scholz},\ and\ \citenamefont {Zeller}}]{dederichs1978lattice}%
  \BibitemOpen
  \bibfield  {author} {\bibinfo {author} {\bibnamefont {Dederichs},
  \bibfnamefont {P.}}, \bibinfo {author} {\bibnamefont {Lehmann}, \bibfnamefont
  {C.}}, \bibinfo {author} {\bibnamefont {Schober}, \bibfnamefont {H.}},
  \bibinfo {author} {\bibnamefont {Scholz}, \bibfnamefont {A.}}, and\ \bibinfo
  {author} {\bibnamefont {Zeller}, \bibfnamefont {R.}},\ }\bibfield  {title}
  {\enquote {\bibinfo {title} {Lattice theory of point defects},}\ }\href@noop
  {} {\bibfield  {journal} {\bibinfo  {journal} {Journal of Nuclear Materials}\
  }\textbf {\bibinfo {volume} {69}},\ \bibinfo {pages} {176--199} (\bibinfo
  {year} {1978})}\BibitemShut {NoStop}%
\bibitem [{\citenamefont {Eggeman}, \citenamefont {Krakow},\ and\ \citenamefont
  {Midgley}(2015)}]{eggeman2015scanning}%
  \BibitemOpen
  \bibfield  {author} {\bibinfo {author} {\bibnamefont {Eggeman}, \bibfnamefont
  {A.~S.}}, \bibinfo {author} {\bibnamefont {Krakow}, \bibfnamefont {R.}}, and\
  \bibinfo {author} {\bibnamefont {Midgley}, \bibfnamefont {P.~A.}},\
  }\bibfield  {title} {\enquote {\bibinfo {title} {Scanning precession electron
  tomography for three-dimensional nanoscale orientation imaging and
  crystallographic analysis},}\ }\href@noop {} {\bibfield  {journal} {\bibinfo
  {journal} {Nature communications}\ }\textbf {\bibinfo {volume} {6}},\
  \bibinfo {pages} {1--7} (\bibinfo {year} {2015})}\BibitemShut {NoStop}%
\bibitem [{\citenamefont {Ewald}(1921)}]{ewald1921berechunung}%
  \BibitemOpen
  \bibfield  {author} {\bibinfo {author} {\bibnamefont {Ewald}, \bibfnamefont
  {P.}},\ }\bibfield  {title} {\enquote {\bibinfo {title} {Die berechunung
  optischer und electrostatischer gitterpotentiale},}\ }\href@noop {}
  {\bibfield  {journal} {\bibinfo  {journal} {Ann. Phys}\ }\textbf {\bibinfo
  {volume} {64}},\ \bibinfo {pages} {253} (\bibinfo {year} {1921})}\BibitemShut
  {NoStop}%
\bibitem [{\citenamefont {Fultz}\ and\ \citenamefont
  {Howe}(2012)}]{fultz2012transmission}%
  \BibitemOpen
  \bibfield  {author} {\bibinfo {author} {\bibnamefont {Fultz}, \bibfnamefont
  {B.}}and\ \bibinfo {author} {\bibnamefont {Howe}, \bibfnamefont {J.~M.}},\
  }\href@noop {} {\emph {\bibinfo {title} {Transmission electron microscopy and
  diffractometry of materials}}}\ (\bibinfo  {publisher} {Springer Science \&
  Business Media},\ \bibinfo {year} {2012})\BibitemShut {NoStop}%
\bibitem [{\citenamefont {Gibbs}(1884)}]{gibbs1884elements}%
  \BibitemOpen
  \bibfield  {author} {\bibinfo {author} {\bibnamefont {Gibbs}, \bibfnamefont
  {J.~W.}},\ }\href@noop {} {\emph {\bibinfo {title} {Elements of vector
  analysis: arranged for the use of students in physics}}}\ (\bibinfo
  {publisher} {Tuttle, Morehouse \& Taylor},\ \bibinfo {year}
  {1884})\BibitemShut {NoStop}%
\bibitem [{\citenamefont {Hong}\ \emph {et~al.}(2021)\citenamefont {Hong},
  \citenamefont {Zeltmann}, \citenamefont {Savitzky}, \citenamefont {DaCosta},
  \citenamefont {M{\"u}ller}, \citenamefont {Minor}, \citenamefont {Bustillo},\
  and\ \citenamefont {Ophus}}]{hong2021multibeam}%
  \BibitemOpen
  \bibfield  {author} {\bibinfo {author} {\bibnamefont {Hong}, \bibfnamefont
  {X.}}, \bibinfo {author} {\bibnamefont {Zeltmann}, \bibfnamefont {S.~E.}},
  \bibinfo {author} {\bibnamefont {Savitzky}, \bibfnamefont {B.~H.}}, \bibinfo
  {author} {\bibnamefont {DaCosta}, \bibfnamefont {L.~R.}}, \bibinfo {author}
  {\bibnamefont {M{\"u}ller}, \bibfnamefont {A.}}, \bibinfo {author}
  {\bibnamefont {Minor}, \bibfnamefont {A.~M.}}, \bibinfo {author}
  {\bibnamefont {Bustillo}, \bibfnamefont {K.~C.}}, and\ \bibinfo {author}
  {\bibnamefont {Ophus}, \bibfnamefont {C.}},\ }\bibfield  {title} {\enquote
  {\bibinfo {title} {Multibeam electron diffraction},}\ }\href@noop {}
  {\bibfield  {journal} {\bibinfo  {journal} {Microscopy and Microanalysis}\
  }\textbf {\bibinfo {volume} {27}},\ \bibinfo {pages} {129--139} (\bibinfo
  {year} {2021})}\BibitemShut {NoStop}%
\bibitem [{\citenamefont {Humphreys}(2001)}]{humphreys2001review}%
  \BibitemOpen
  \bibfield  {author} {\bibinfo {author} {\bibnamefont {Humphreys},
  \bibfnamefont {F.}},\ }\bibfield  {title} {\enquote {\bibinfo {title} {Review
  grain and subgrain characterisation by electron backscatter diffraction},}\
  }\href@noop {} {\bibfield  {journal} {\bibinfo  {journal} {Journal of
  materials science}\ }\textbf {\bibinfo {volume} {36}},\ \bibinfo {pages}
  {3833--3854} (\bibinfo {year} {2001})}\BibitemShut {NoStop}%
\bibitem [{\citenamefont {Jain}\ \emph {et~al.}(2013)\citenamefont {Jain},
  \citenamefont {Ong}, \citenamefont {Hautier}, \citenamefont {Chen},
  \citenamefont {Richards}, \citenamefont {Dacek}, \citenamefont {Cholia},
  \citenamefont {Gunter}, \citenamefont {Skinner}, \citenamefont {Ceder} \emph
  {et~al.}}]{jain2013commentary}%
  \BibitemOpen
  \bibfield  {author} {\bibinfo {author} {\bibnamefont {Jain}, \bibfnamefont
  {A.}}, \bibinfo {author} {\bibnamefont {Ong}, \bibfnamefont {S.~P.}},
  \bibinfo {author} {\bibnamefont {Hautier}, \bibfnamefont {G.}}, \bibinfo
  {author} {\bibnamefont {Chen}, \bibfnamefont {W.}}, \bibinfo {author}
  {\bibnamefont {Richards}, \bibfnamefont {W.~D.}}, \bibinfo {author}
  {\bibnamefont {Dacek}, \bibfnamefont {S.}}, \bibinfo {author} {\bibnamefont
  {Cholia}, \bibfnamefont {S.}}, \bibinfo {author} {\bibnamefont {Gunter},
  \bibfnamefont {D.}}, \bibinfo {author} {\bibnamefont {Skinner}, \bibfnamefont
  {D.}}, \bibinfo {author} {\bibnamefont {Ceder}, \bibfnamefont {G.}},  \emph
  {et~al.},\ }\bibfield  {title} {\enquote {\bibinfo {title} {Commentary: The
  materials project: A materials genome approach to accelerating materials
  innovation},}\ }\href@noop {} {\bibfield  {journal} {\bibinfo  {journal} {APL
  materials}\ }\textbf {\bibinfo {volume} {1}},\ \bibinfo {pages} {011002}
  (\bibinfo {year} {2013})}\BibitemShut {NoStop}%
\bibitem [{\citenamefont {Janssen}(2007)}]{janssen2007stress}%
  \BibitemOpen
  \bibfield  {author} {\bibinfo {author} {\bibnamefont {Janssen}, \bibfnamefont
  {G.}},\ }\bibfield  {title} {\enquote {\bibinfo {title} {Stress and strain in
  polycrystalline thin films},}\ }\href@noop {} {\bibfield  {journal} {\bibinfo
   {journal} {Thin solid films}\ }\textbf {\bibinfo {volume} {515}},\ \bibinfo
  {pages} {6654--6664} (\bibinfo {year} {2007})}\BibitemShut {NoStop}%
\bibitem [{\citenamefont {Jeong}\ \emph {et~al.}(2021)\citenamefont {Jeong},
  \citenamefont {Cautaerts}, \citenamefont {Dehm},\ and\ \citenamefont
  {Liebscher}}]{jeong2021automated}%
  \BibitemOpen
  \bibfield  {author} {\bibinfo {author} {\bibnamefont {Jeong}, \bibfnamefont
  {J.}}, \bibinfo {author} {\bibnamefont {Cautaerts}, \bibfnamefont {N.}},
  \bibinfo {author} {\bibnamefont {Dehm}, \bibfnamefont {G.}}, and\ \bibinfo
  {author} {\bibnamefont {Liebscher}, \bibfnamefont {C.~H.}},\ }\bibfield
  {title} {\enquote {\bibinfo {title} {Automated crystal orientation mapping by
  precession electron diffraction assisted four-dimensional scanning
  transmission electron microscopy ({4D-STEM}) using a scintillator based cmos
  detector},}\ }\href@noop {} {\bibfield  {journal} {\bibinfo  {journal} {arXiv
  preprint arXiv:2102.09711}\ } (\bibinfo {year} {2021})}\BibitemShut {NoStop}%
\bibitem [{\citenamefont {Kirkland}(2020)}]{kirkland2020advanced}%
  \BibitemOpen
  \bibfield  {author} {\bibinfo {author} {\bibnamefont {Kirkland},
  \bibfnamefont {E.~J.}},\ }\href@noop {} {\emph {\bibinfo {title} {Advanced
  computing in electron microscopy, 3rd edition}}}\ (\bibinfo  {publisher}
  {Springer Science \& Business Media},\ \bibinfo {year} {2020})\BibitemShut
  {NoStop}%
\bibitem [{\citenamefont {Kobler}\ \emph {et~al.}(2013)\citenamefont {Kobler},
  \citenamefont {Kashiwar}, \citenamefont {Hahn},\ and\ \citenamefont
  {K{\"u}bel}}]{kobler2013combination}%
  \BibitemOpen
  \bibfield  {author} {\bibinfo {author} {\bibnamefont {Kobler}, \bibfnamefont
  {A.}}, \bibinfo {author} {\bibnamefont {Kashiwar}, \bibfnamefont {A.}},
  \bibinfo {author} {\bibnamefont {Hahn}, \bibfnamefont {H.}}, and\ \bibinfo
  {author} {\bibnamefont {K{\"u}bel}, \bibfnamefont {C.}},\ }\bibfield  {title}
  {\enquote {\bibinfo {title} {Combination of in situ straining and {ACOM TEM}:
  A novel method for analysis of plastic deformation of nanocrystalline
  metals},}\ }\href@noop {} {\bibfield  {journal} {\bibinfo  {journal}
  {Ultramicroscopy}\ }\textbf {\bibinfo {volume} {128}},\ \bibinfo {pages}
  {68--81} (\bibinfo {year} {2013})}\BibitemShut {NoStop}%
\bibitem [{\citenamefont {Lang}\ \emph {et~al.}(2021)\citenamefont {Lang},
  \citenamefont {Taylor}, \citenamefont {Pratt}, \citenamefont {Nenoff},\ and\
  \citenamefont {Hattar}}]{lang2021automated}%
  \BibitemOpen
  \bibfield  {author} {\bibinfo {author} {\bibnamefont {Lang}, \bibfnamefont
  {E.}}, \bibinfo {author} {\bibnamefont {Taylor}, \bibfnamefont {C.}},
  \bibinfo {author} {\bibnamefont {Pratt}, \bibfnamefont {S.}}, \bibinfo
  {author} {\bibnamefont {Nenoff}, \bibfnamefont {T.}}, and\ \bibinfo {author}
  {\bibnamefont {Hattar}, \bibfnamefont {K.}},\ }\bibfield  {title} {\enquote
  {\bibinfo {title} {Automated crystal orientation mapping with a liquid-cell
  tem},}\ }\href@noop {} {\bibfield  {journal} {\bibinfo  {journal} {Microscopy
  and Microanalysis}\ }\textbf {\bibinfo {volume} {27}},\ \bibinfo {pages}
  {2232--2233} (\bibinfo {year} {2021})}\BibitemShut {NoStop}%
\bibitem [{\citenamefont {LeSar}(2014)}]{lesar2014simulations}%
  \BibitemOpen
  \bibfield  {author} {\bibinfo {author} {\bibnamefont {LeSar}, \bibfnamefont
  {R.}},\ }\bibfield  {title} {\enquote {\bibinfo {title} {Simulations of
  dislocation structure and response},}\ }\href@noop {} {\bibfield  {journal}
  {\bibinfo  {journal} {Annu. Rev. Condens. Matter Phys.}\ }\textbf {\bibinfo
  {volume} {5}},\ \bibinfo {pages} {375--407} (\bibinfo {year}
  {2014})}\BibitemShut {NoStop}%
\bibitem [{\citenamefont {Li}\ \emph {et~al.}(2020)\citenamefont {Li},
  \citenamefont {Jin}, \citenamefont {Zhou},\ and\ \citenamefont
  {Lu}}]{li2020constrained}%
  \BibitemOpen
  \bibfield  {author} {\bibinfo {author} {\bibnamefont {Li}, \bibfnamefont
  {X.}}, \bibinfo {author} {\bibnamefont {Jin}, \bibfnamefont {Z.}}, \bibinfo
  {author} {\bibnamefont {Zhou}, \bibfnamefont {X.}}, and\ \bibinfo {author}
  {\bibnamefont {Lu}, \bibfnamefont {K.}},\ }\bibfield  {title} {\enquote
  {\bibinfo {title} {Constrained minimal-interface structures in
  polycrystalline copper with extremely fine grains},}\ }\href@noop {}
  {\bibfield  {journal} {\bibinfo  {journal} {Science}\ }\textbf {\bibinfo
  {volume} {370}},\ \bibinfo {pages} {831--836} (\bibinfo {year}
  {2020})}\BibitemShut {NoStop}%
\bibitem [{\citenamefont {Linck}\ \emph {et~al.}(2016)\citenamefont {Linck},
  \citenamefont {Hartel}, \citenamefont {Uhlemann}, \citenamefont {Kahl},
  \citenamefont {M{\"u}ller}, \citenamefont {Zach}, \citenamefont {Haider},
  \citenamefont {Niestadt}, \citenamefont {Bischoff}, \citenamefont {Biskupek}
  \emph {et~al.}}]{linck2016chromatic}%
  \BibitemOpen
  \bibfield  {author} {\bibinfo {author} {\bibnamefont {Linck}, \bibfnamefont
  {M.}}, \bibinfo {author} {\bibnamefont {Hartel}, \bibfnamefont {P.}},
  \bibinfo {author} {\bibnamefont {Uhlemann}, \bibfnamefont {S.}}, \bibinfo
  {author} {\bibnamefont {Kahl}, \bibfnamefont {F.}}, \bibinfo {author}
  {\bibnamefont {M{\"u}ller}, \bibfnamefont {H.}}, \bibinfo {author}
  {\bibnamefont {Zach}, \bibfnamefont {J.}}, \bibinfo {author} {\bibnamefont
  {Haider}, \bibfnamefont {M.}}, \bibinfo {author} {\bibnamefont {Niestadt},
  \bibfnamefont {M.}}, \bibinfo {author} {\bibnamefont {Bischoff},
  \bibfnamefont {M.}}, \bibinfo {author} {\bibnamefont {Biskupek},
  \bibfnamefont {J.}},  \emph {et~al.},\ }\bibfield  {title} {\enquote
  {\bibinfo {title} {Chromatic aberration correction for atomic resolution
  {TEM} imaging from 20 to 80 kv},}\ }\href@noop {} {\bibfield  {journal}
  {\bibinfo  {journal} {Physical review letters}\ }\textbf {\bibinfo {volume}
  {117}},\ \bibinfo {pages} {076101} (\bibinfo {year} {2016})}\BibitemShut
  {NoStop}%
\bibitem [{\citenamefont {Lobato}\ and\ \citenamefont
  {Van~Dyck}(2014)}]{lobato2014accurate}%
  \BibitemOpen
  \bibfield  {author} {\bibinfo {author} {\bibnamefont {Lobato}, \bibfnamefont
  {I.}}and\ \bibinfo {author} {\bibnamefont {Van~Dyck}, \bibfnamefont {D.}},\
  }\bibfield  {title} {\enquote {\bibinfo {title} {An accurate parameterization
  for scattering factors, electron densities and electrostatic potentials for
  neutral atoms that obey all physical constraints},}\ }\href@noop {}
  {\bibfield  {journal} {\bibinfo  {journal} {Acta Crystallographica Section A:
  Foundations and Advances}\ }\textbf {\bibinfo {volume} {70}},\ \bibinfo
  {pages} {636--649} (\bibinfo {year} {2014})}\BibitemShut {NoStop}%
\bibitem [{\citenamefont {Londo{\~n}o-Calderon}\ \emph
  {et~al.}(2020)\citenamefont {Londo{\~n}o-Calderon}, \citenamefont {Williams},
  \citenamefont {Ophus},\ and\ \citenamefont {Pettes}}]{londono20201d}%
  \BibitemOpen
  \bibfield  {author} {\bibinfo {author} {\bibnamefont {Londo{\~n}o-Calderon},
  \bibfnamefont {A.}}, \bibinfo {author} {\bibnamefont {Williams},
  \bibfnamefont {D.~J.}}, \bibinfo {author} {\bibnamefont {Ophus},
  \bibfnamefont {C.}}, and\ \bibinfo {author} {\bibnamefont {Pettes},
  \bibfnamefont {M.~T.}},\ }\bibfield  {title} {\enquote {\bibinfo {title} {1d
  to 2d transition in tellurium observed by 4d electron microscopy},}\
  }\href@noop {} {\bibfield  {journal} {\bibinfo  {journal} {Small}\ }\textbf
  {\bibinfo {volume} {16}},\ \bibinfo {pages} {2005447} (\bibinfo {year}
  {2020})}\BibitemShut {NoStop}%
\bibitem [{\citenamefont {Londo{\~n}o-Calderon}\ \emph
  {et~al.}(2021)\citenamefont {Londo{\~n}o-Calderon}, \citenamefont {Williams},
  \citenamefont {Schneider}, \citenamefont {Savitzky}, \citenamefont {Ophus},
  \citenamefont {Ma}, \citenamefont {Zhu},\ and\ \citenamefont
  {Pettes}}]{londono2021intrinsic}%
  \BibitemOpen
  \bibfield  {author} {\bibinfo {author} {\bibnamefont {Londo{\~n}o-Calderon},
  \bibfnamefont {A.}}, \bibinfo {author} {\bibnamefont {Williams},
  \bibfnamefont {D.~J.}}, \bibinfo {author} {\bibnamefont {Schneider},
  \bibfnamefont {M.~M.}}, \bibinfo {author} {\bibnamefont {Savitzky},
  \bibfnamefont {B.~H.}}, \bibinfo {author} {\bibnamefont {Ophus},
  \bibfnamefont {C.}}, \bibinfo {author} {\bibnamefont {Ma}, \bibfnamefont
  {S.}}, \bibinfo {author} {\bibnamefont {Zhu}, \bibfnamefont {H.}}, and\
  \bibinfo {author} {\bibnamefont {Pettes}, \bibfnamefont {M.~T.}},\ }\bibfield
   {title} {\enquote {\bibinfo {title} {Intrinsic helical twist and chirality
  in ultrathin tellurium nanowires},}\ }\href@noop {} {\bibfield  {journal}
  {\bibinfo  {journal} {Nanoscale}\ }\textbf {\bibinfo {volume} {13}},\
  \bibinfo {pages} {9606--9614} (\bibinfo {year} {2021})}\BibitemShut {NoStop}%
\bibitem [{\citenamefont {MacLaren}\ \emph {et~al.}(2020)\citenamefont
  {MacLaren}, \citenamefont {Frutos-Myro}, \citenamefont {McGrouther},
  \citenamefont {McFadzean}, \citenamefont {Weiss}, \citenamefont {Cosart},
  \citenamefont {Portillo}, \citenamefont {Robins}, \citenamefont
  {Nicolopoulos}, \citenamefont {Del~Busto} \emph
  {et~al.}}]{maclaren2020comparison}%
  \BibitemOpen
  \bibfield  {author} {\bibinfo {author} {\bibnamefont {MacLaren},
  \bibfnamefont {I.}}, \bibinfo {author} {\bibnamefont {Frutos-Myro},
  \bibfnamefont {E.}}, \bibinfo {author} {\bibnamefont {McGrouther},
  \bibfnamefont {D.}}, \bibinfo {author} {\bibnamefont {McFadzean},
  \bibfnamefont {S.}}, \bibinfo {author} {\bibnamefont {Weiss}, \bibfnamefont
  {J.~K.}}, \bibinfo {author} {\bibnamefont {Cosart}, \bibfnamefont {D.}},
  \bibinfo {author} {\bibnamefont {Portillo}, \bibfnamefont {J.}}, \bibinfo
  {author} {\bibnamefont {Robins}, \bibfnamefont {A.}}, \bibinfo {author}
  {\bibnamefont {Nicolopoulos}, \bibfnamefont {S.}}, \bibinfo {author}
  {\bibnamefont {Del~Busto}, \bibfnamefont {E.~N.}},  \emph {et~al.},\
  }\bibfield  {title} {\enquote {\bibinfo {title} {A comparison of a direct
  electron detector and a high-speed video camera for a scanning precession
  electron diffraction phase and orientation mapping},}\ }\href@noop {}
  {\bibfield  {journal} {\bibinfo  {journal} {Microscopy and Microanalysis}\
  }\textbf {\bibinfo {volume} {26}},\ \bibinfo {pages} {1110--1116} (\bibinfo
  {year} {2020})}\BibitemShut {NoStop}%
\bibitem [{\citenamefont {Maeland}\ and\ \citenamefont
  {Flanagan}(1964)}]{maeland1964lattice}%
  \BibitemOpen
  \bibfield  {author} {\bibinfo {author} {\bibnamefont {Maeland}, \bibfnamefont
  {A.}}and\ \bibinfo {author} {\bibnamefont {Flanagan}, \bibfnamefont
  {T.~B.}},\ }\bibfield  {title} {\enquote {\bibinfo {title} {Lattice spacings
  of gold--palladium alloys},}\ }\href@noop {} {\bibfield  {journal} {\bibinfo
  {journal} {Canadian journal of physics}\ }\textbf {\bibinfo {volume} {42}},\
  \bibinfo {pages} {2364--2366} (\bibinfo {year} {1964})}\BibitemShut {NoStop}%
\bibitem [{\citenamefont {Mehta}\ \emph {et~al.}(2020)\citenamefont {Mehta},
  \citenamefont {Gauquelin}, \citenamefont {Nord}, \citenamefont {Orekhov},
  \citenamefont {Bender}, \citenamefont {Cerbu}, \citenamefont {Verbeeck},\
  and\ \citenamefont {Vandervorst}}]{mehta2020unravelling}%
  \BibitemOpen
  \bibfield  {author} {\bibinfo {author} {\bibnamefont {Mehta}, \bibfnamefont
  {A.~N.}}, \bibinfo {author} {\bibnamefont {Gauquelin}, \bibfnamefont {N.}},
  \bibinfo {author} {\bibnamefont {Nord}, \bibfnamefont {M.}}, \bibinfo
  {author} {\bibnamefont {Orekhov}, \bibfnamefont {A.}}, \bibinfo {author}
  {\bibnamefont {Bender}, \bibfnamefont {H.}}, \bibinfo {author} {\bibnamefont
  {Cerbu}, \bibfnamefont {D.}}, \bibinfo {author} {\bibnamefont {Verbeeck},
  \bibfnamefont {J.}}, and\ \bibinfo {author} {\bibnamefont {Vandervorst},
  \bibfnamefont {W.}},\ }\bibfield  {title} {\enquote {\bibinfo {title}
  {Unravelling stacking order in epitaxial bilayer mx2 using {4D-STEM} with
  unsupervised learning},}\ }\href@noop {} {\bibfield  {journal} {\bibinfo
  {journal} {Nanotechnology}\ }\textbf {\bibinfo {volume} {31}},\ \bibinfo
  {pages} {445702} (\bibinfo {year} {2020})}\BibitemShut {NoStop}%
\bibitem [{\citenamefont {Meng}\ and\ \citenamefont
  {Zuo}(2016)}]{meng2016three}%
  \BibitemOpen
  \bibfield  {author} {\bibinfo {author} {\bibnamefont {Meng}, \bibfnamefont
  {Y.}}and\ \bibinfo {author} {\bibnamefont {Zuo}, \bibfnamefont {J.-M.}},\
  }\bibfield  {title} {\enquote {\bibinfo {title} {Three-dimensional
  nanostructure determination from a large diffraction data set recorded using
  scanning electron nanodiffraction},}\ }\href@noop {} {\bibfield  {journal}
  {\bibinfo  {journal} {IUCrJ}\ }\textbf {\bibinfo {volume} {3}},\ \bibinfo
  {pages} {300--308} (\bibinfo {year} {2016})}\BibitemShut {NoStop}%
\bibitem [{\citenamefont {Midgley}\ and\ \citenamefont
  {Eggeman}(2015)}]{midgley2015precession}%
  \BibitemOpen
  \bibfield  {author} {\bibinfo {author} {\bibnamefont {Midgley}, \bibfnamefont
  {P.~A.}}and\ \bibinfo {author} {\bibnamefont {Eggeman}, \bibfnamefont
  {A.~S.}},\ }\bibfield  {title} {\enquote {\bibinfo {title} {Precession
  electron diffraction--a topical review},}\ }\href@noop {} {\bibfield
  {journal} {\bibinfo  {journal} {IUCrJ}\ }\textbf {\bibinfo {volume} {2}},\
  \bibinfo {pages} {126--136} (\bibinfo {year} {2015})}\BibitemShut {NoStop}%
\bibitem [{\citenamefont {Moeck}\ \emph {et~al.}(2011)\citenamefont {Moeck},
  \citenamefont {Rouvimov}, \citenamefont {Rauch}, \citenamefont {V{\'e}ron},
  \citenamefont {Kirmse}, \citenamefont {H{\"a}usler}, \citenamefont {Neumann},
  \citenamefont {Bultreys}, \citenamefont {Maniette},\ and\ \citenamefont
  {Nicolopoulos}}]{moeck2011high}%
  \BibitemOpen
  \bibfield  {author} {\bibinfo {author} {\bibnamefont {Moeck}, \bibfnamefont
  {P.}}, \bibinfo {author} {\bibnamefont {Rouvimov}, \bibfnamefont {S.}},
  \bibinfo {author} {\bibnamefont {Rauch}, \bibfnamefont {E.}}, \bibinfo
  {author} {\bibnamefont {V{\'e}ron}, \bibfnamefont {M.}}, \bibinfo {author}
  {\bibnamefont {Kirmse}, \bibfnamefont {H.}}, \bibinfo {author} {\bibnamefont
  {H{\"a}usler}, \bibfnamefont {I.}}, \bibinfo {author} {\bibnamefont
  {Neumann}, \bibfnamefont {W.}}, \bibinfo {author} {\bibnamefont {Bultreys},
  \bibfnamefont {D.}}, \bibinfo {author} {\bibnamefont {Maniette},
  \bibfnamefont {Y.}}, and\ \bibinfo {author} {\bibnamefont {Nicolopoulos},
  \bibfnamefont {S.}},\ }\bibfield  {title} {\enquote {\bibinfo {title} {High
  spatial resolution semi-automatic crystallite orientation and phase mapping
  of nanocrystals in transmission electron microscopes},}\ }\href@noop {}
  {\bibfield  {journal} {\bibinfo  {journal} {Crystal research and technology}\
  }\textbf {\bibinfo {volume} {46}},\ \bibinfo {pages} {589--606} (\bibinfo
  {year} {2011})}\BibitemShut {NoStop}%
\bibitem [{\citenamefont {Munshi}\ \emph {et~al.}(2021)\citenamefont {Munshi},
  \citenamefont {Rakowski}, \citenamefont {Savitzky}, \citenamefont {Zeltmann},
  \citenamefont {Henderson}, \citenamefont {Cholia}, \citenamefont {Chan},\
  and\ \citenamefont {Ophus}}]{munshi2021ml}%
  \BibitemOpen
  \bibfield  {author} {\bibinfo {author} {\bibnamefont {Munshi}, \bibfnamefont
  {J.}}, \bibinfo {author} {\bibnamefont {Rakowski}, \bibfnamefont {A.}},
  \bibinfo {author} {\bibnamefont {Savitzky}, \bibfnamefont {B.~H.}}, \bibinfo
  {author} {\bibnamefont {Zeltmann}, \bibfnamefont {S.~E.}}, \bibinfo {author}
  {\bibnamefont {Henderson}, \bibfnamefont {M.}}, \bibinfo {author}
  {\bibnamefont {Cholia}, \bibfnamefont {S.}}, \bibinfo {author} {\bibnamefont
  {Chan}, \bibfnamefont {M.}}, and\ \bibinfo {author} {\bibnamefont {Ophus},
  \bibfnamefont {C.}},\ }\bibfield  {title} {\enquote {\bibinfo {title}
  {Improving strain mapping from electron diffraction patterns using fourier
  spacedeep learning networks},}\ }\href@noop {} {\bibfield  {journal}
  {\bibinfo  {journal} {manuscript in preparation}\ } (\bibinfo {year}
  {2021})}\BibitemShut {NoStop}%
\bibitem [{\citenamefont {Nord}\ \emph {et~al.}(2020)\citenamefont {Nord},
  \citenamefont {Webster}, \citenamefont {Paton}, \citenamefont {McVitie},
  \citenamefont {McGrouther}, \citenamefont {MacLaren},\ and\ \citenamefont
  {Paterson}}]{nord2020fast}%
  \BibitemOpen
  \bibfield  {author} {\bibinfo {author} {\bibnamefont {Nord}, \bibfnamefont
  {M.}}, \bibinfo {author} {\bibnamefont {Webster}, \bibfnamefont {R.~W.}},
  \bibinfo {author} {\bibnamefont {Paton}, \bibfnamefont {K.~A.}}, \bibinfo
  {author} {\bibnamefont {McVitie}, \bibfnamefont {S.}}, \bibinfo {author}
  {\bibnamefont {McGrouther}, \bibfnamefont {D.}}, \bibinfo {author}
  {\bibnamefont {MacLaren}, \bibfnamefont {I.}}, and\ \bibinfo {author}
  {\bibnamefont {Paterson}, \bibfnamefont {G.~W.}},\ }\bibfield  {title}
  {\enquote {\bibinfo {title} {Fast pixelated detectors in scanning
  transmission electron microscopy. part i: data acquisition, live processing,
  and storage},}\ }\href@noop {} {\bibfield  {journal} {\bibinfo  {journal}
  {Microscopy and Microanalysis}\ }\textbf {\bibinfo {volume} {26}},\ \bibinfo
  {pages} {653--666} (\bibinfo {year} {2020})}\BibitemShut {NoStop}%
\bibitem [{\citenamefont {Ong}\ \emph {et~al.}(2013)\citenamefont {Ong},
  \citenamefont {Richards}, \citenamefont {Jain}, \citenamefont {Hautier},
  \citenamefont {Kocher}, \citenamefont {Cholia}, \citenamefont {Gunter},
  \citenamefont {Chevrier}, \citenamefont {Persson},\ and\ \citenamefont
  {Ceder}}]{ong2013python}%
  \BibitemOpen
  \bibfield  {author} {\bibinfo {author} {\bibnamefont {Ong}, \bibfnamefont
  {S.~P.}}, \bibinfo {author} {\bibnamefont {Richards}, \bibfnamefont {W.~D.}},
  \bibinfo {author} {\bibnamefont {Jain}, \bibfnamefont {A.}}, \bibinfo
  {author} {\bibnamefont {Hautier}, \bibfnamefont {G.}}, \bibinfo {author}
  {\bibnamefont {Kocher}, \bibfnamefont {M.}}, \bibinfo {author} {\bibnamefont
  {Cholia}, \bibfnamefont {S.}}, \bibinfo {author} {\bibnamefont {Gunter},
  \bibfnamefont {D.}}, \bibinfo {author} {\bibnamefont {Chevrier},
  \bibfnamefont {V.~L.}}, \bibinfo {author} {\bibnamefont {Persson},
  \bibfnamefont {K.~A.}}, and\ \bibinfo {author} {\bibnamefont {Ceder},
  \bibfnamefont {G.}},\ }\bibfield  {title} {\enquote {\bibinfo {title} {Python
  materials genomics (pymatgen): A robust, open-source python library for
  materials analysis},}\ }\href@noop {} {\bibfield  {journal} {\bibinfo
  {journal} {Computational Materials Science}\ }\textbf {\bibinfo {volume}
  {68}},\ \bibinfo {pages} {314--319} (\bibinfo {year} {2013})}\BibitemShut
  {NoStop}%
\bibitem [{\citenamefont {Ophus}(2017)}]{ophus2017fast}%
  \BibitemOpen
  \bibfield  {author} {\bibinfo {author} {\bibnamefont {Ophus}, \bibfnamefont
  {C.}},\ }\bibfield  {title} {\enquote {\bibinfo {title} {A fast image
  simulation algorithm for scanning transmission electron microscopy},}\
  }\href@noop {} {\bibfield  {journal} {\bibinfo  {journal} {Advanced
  structural and chemical imaging}\ }\textbf {\bibinfo {volume} {3}},\ \bibinfo
  {pages} {1--11} (\bibinfo {year} {2017})}\BibitemShut {NoStop}%
\bibitem [{\citenamefont {Ophus}(2019)}]{ophus2019four}%
  \BibitemOpen
  \bibfield  {author} {\bibinfo {author} {\bibnamefont {Ophus}, \bibfnamefont
  {C.}},\ }\bibfield  {title} {\enquote {\bibinfo {title} {{Four-dimensional
  scanning transmission electron microscopy ({4D-STEM}): From scanning
  nanodiffraction to ptychography and beyond}},}\ }\href@noop {} {\bibfield
  {journal} {\bibinfo  {journal} {Microscopy and Microanalysis}\ }\textbf
  {\bibinfo {volume} {25}},\ \bibinfo {pages} {563--582} (\bibinfo {year}
  {2019})}\BibitemShut {NoStop}%
\bibitem [{\citenamefont {Ophus}\ \emph {et~al.}(2015)\citenamefont {Ophus},
  \citenamefont {Shekhawat}, \citenamefont {Rasool},\ and\ \citenamefont
  {Zettl}}]{ophus2015large}%
  \BibitemOpen
  \bibfield  {author} {\bibinfo {author} {\bibnamefont {Ophus}, \bibfnamefont
  {C.}}, \bibinfo {author} {\bibnamefont {Shekhawat}, \bibfnamefont {A.}},
  \bibinfo {author} {\bibnamefont {Rasool}, \bibfnamefont {H.}}, and\ \bibinfo
  {author} {\bibnamefont {Zettl}, \bibfnamefont {A.}},\ }\bibfield  {title}
  {\enquote {\bibinfo {title} {Large-scale experimental and theoretical study
  of graphene grain boundary structures},}\ }\href@noop {} {\bibfield
  {journal} {\bibinfo  {journal} {Physical Review B}\ }\textbf {\bibinfo
  {volume} {92}},\ \bibinfo {pages} {205402} (\bibinfo {year}
  {2015})}\BibitemShut {NoStop}%
\bibitem [{\citenamefont {Ozdol}\ \emph {et~al.}(2015)\citenamefont {Ozdol},
  \citenamefont {Gammer}, \citenamefont {Jin}, \citenamefont {Ercius},
  \citenamefont {Ophus}, \citenamefont {Ciston},\ and\ \citenamefont
  {Minor}}]{ozdol2015strain}%
  \BibitemOpen
  \bibfield  {author} {\bibinfo {author} {\bibnamefont {Ozdol}, \bibfnamefont
  {V.}}, \bibinfo {author} {\bibnamefont {Gammer}, \bibfnamefont {C.}},
  \bibinfo {author} {\bibnamefont {Jin}, \bibfnamefont {X.}}, \bibinfo {author}
  {\bibnamefont {Ercius}, \bibfnamefont {P.}}, \bibinfo {author} {\bibnamefont
  {Ophus}, \bibfnamefont {C.}}, \bibinfo {author} {\bibnamefont {Ciston},
  \bibfnamefont {J.}}, and\ \bibinfo {author} {\bibnamefont {Minor},
  \bibfnamefont {A.}},\ }\bibfield  {title} {\enquote {\bibinfo {title} {Strain
  mapping at nanometer resolution using advanced nano-beam electron
  diffraction},}\ }\href@noop {} {\bibfield  {journal} {\bibinfo  {journal}
  {Applied Physics Letters}\ }\textbf {\bibinfo {volume} {106}},\ \bibinfo
  {pages} {253107} (\bibinfo {year} {2015})}\BibitemShut {NoStop}%
\bibitem [{\citenamefont {Park}\ \emph {et~al.}(2019)\citenamefont {Park},
  \citenamefont {Park}, \citenamefont {Lee}, \citenamefont {So}, \citenamefont
  {Kim}, \citenamefont {Jeong}, \citenamefont {Han}, \citenamefont {Wolf},
  \citenamefont {Lee}, \citenamefont {Yoo} \emph {et~al.}}]{park2019efficient}%
  \BibitemOpen
  \bibfield  {author} {\bibinfo {author} {\bibnamefont {Park}, \bibfnamefont
  {M.-H.}}, \bibinfo {author} {\bibnamefont {Park}, \bibfnamefont {J.}},
  \bibinfo {author} {\bibnamefont {Lee}, \bibfnamefont {J.}}, \bibinfo {author}
  {\bibnamefont {So}, \bibfnamefont {H.~S.}}, \bibinfo {author} {\bibnamefont
  {Kim}, \bibfnamefont {H.}}, \bibinfo {author} {\bibnamefont {Jeong},
  \bibfnamefont {S.-H.}}, \bibinfo {author} {\bibnamefont {Han}, \bibfnamefont
  {T.-H.}}, \bibinfo {author} {\bibnamefont {Wolf}, \bibfnamefont {C.}},
  \bibinfo {author} {\bibnamefont {Lee}, \bibfnamefont {H.}}, \bibinfo {author}
  {\bibnamefont {Yoo}, \bibfnamefont {S.}},  \emph {et~al.},\ }\bibfield
  {title} {\enquote {\bibinfo {title} {Efficient perovskite light-emitting
  diodes using polycrystalline core--shell-mimicked nanograins},}\ }\href@noop
  {} {\bibfield  {journal} {\bibinfo  {journal} {Advanced Functional
  Materials}\ }\textbf {\bibinfo {volume} {29}},\ \bibinfo {pages} {1902017}
  (\bibinfo {year} {2019})}\BibitemShut {NoStop}%
\bibitem [{\citenamefont {Paterson}\ \emph {et~al.}(2020)\citenamefont
  {Paterson}, \citenamefont {Webster}, \citenamefont {Ross}, \citenamefont
  {Paton}, \citenamefont {Macgregor}, \citenamefont {McGrouther}, \citenamefont
  {MacLaren},\ and\ \citenamefont {Nord}}]{paterson2020fast}%
  \BibitemOpen
  \bibfield  {author} {\bibinfo {author} {\bibnamefont {Paterson},
  \bibfnamefont {G.~W.}}, \bibinfo {author} {\bibnamefont {Webster},
  \bibfnamefont {R.~W.}}, \bibinfo {author} {\bibnamefont {Ross}, \bibfnamefont
  {A.}}, \bibinfo {author} {\bibnamefont {Paton}, \bibfnamefont {K.~A.}},
  \bibinfo {author} {\bibnamefont {Macgregor}, \bibfnamefont {T.~A.}}, \bibinfo
  {author} {\bibnamefont {McGrouther}, \bibfnamefont {D.}}, \bibinfo {author}
  {\bibnamefont {MacLaren}, \bibfnamefont {I.}}, and\ \bibinfo {author}
  {\bibnamefont {Nord}, \bibfnamefont {M.}},\ }\bibfield  {title} {\enquote
  {\bibinfo {title} {Fast pixelated detectors in scanning transmission electron
  microscopy. part ii: Post-acquisition data processing, visualization, and
  structural characterization},}\ }\href@noop {} {\bibfield  {journal}
  {\bibinfo  {journal} {Microscopy and Microanalysis}\ }\textbf {\bibinfo
  {volume} {26}},\ \bibinfo {pages} {944--963} (\bibinfo {year}
  {2020})}\BibitemShut {NoStop}%
\bibitem [{\citenamefont {Pekin}\ \emph {et~al.}(2017)\citenamefont {Pekin},
  \citenamefont {Gammer}, \citenamefont {Ciston}, \citenamefont {Minor},\ and\
  \citenamefont {Ophus}}]{pekin2017optimizing}%
  \BibitemOpen
  \bibfield  {author} {\bibinfo {author} {\bibnamefont {Pekin}, \bibfnamefont
  {T.~C.}}, \bibinfo {author} {\bibnamefont {Gammer}, \bibfnamefont {C.}},
  \bibinfo {author} {\bibnamefont {Ciston}, \bibfnamefont {J.}}, \bibinfo
  {author} {\bibnamefont {Minor}, \bibfnamefont {A.~M.}}, and\ \bibinfo
  {author} {\bibnamefont {Ophus}, \bibfnamefont {C.}},\ }\bibfield  {title}
  {\enquote {\bibinfo {title} {Optimizing disk registration algorithms for
  nanobeam electron diffraction strain mapping},}\ }\href@noop {} {\bibfield
  {journal} {\bibinfo  {journal} {Ultramicroscopy}\ }\textbf {\bibinfo {volume}
  {176}},\ \bibinfo {pages} {170--176} (\bibinfo {year} {2017})}\BibitemShut
  {NoStop}%
\bibitem [{\citenamefont {Peter}\ \emph {et~al.}(2018)\citenamefont {Peter},
  \citenamefont {Frolov}, \citenamefont {Duarte}, \citenamefont {Hadian},
  \citenamefont {Ophus}, \citenamefont {Kirchlechner}, \citenamefont
  {Liebscher},\ and\ \citenamefont {Dehm}}]{peter2018segregation}%
  \BibitemOpen
  \bibfield  {author} {\bibinfo {author} {\bibnamefont {Peter}, \bibfnamefont
  {N.~J.}}, \bibinfo {author} {\bibnamefont {Frolov}, \bibfnamefont {T.}},
  \bibinfo {author} {\bibnamefont {Duarte}, \bibfnamefont {M.~J.}}, \bibinfo
  {author} {\bibnamefont {Hadian}, \bibfnamefont {R.}}, \bibinfo {author}
  {\bibnamefont {Ophus}, \bibfnamefont {C.}}, \bibinfo {author} {\bibnamefont
  {Kirchlechner}, \bibfnamefont {C.}}, \bibinfo {author} {\bibnamefont
  {Liebscher}, \bibfnamefont {C.~H.}}, and\ \bibinfo {author} {\bibnamefont
  {Dehm}, \bibfnamefont {G.}},\ }\bibfield  {title} {\enquote {\bibinfo {title}
  {Segregation-induced nanofaceting transition at an asymmetric tilt grain
  boundary in copper},}\ }\href@noop {} {\bibfield  {journal} {\bibinfo
  {journal} {Physical review letters}\ }\textbf {\bibinfo {volume} {121}},\
  \bibinfo {pages} {255502} (\bibinfo {year} {2018})}\BibitemShut {NoStop}%
\bibitem [{\citenamefont {Qi}\ \emph {et~al.}(2020)\citenamefont {Qi},
  \citenamefont {Sahabudeen}, \citenamefont {Liang}, \citenamefont
  {Polo{\v{z}}ij}, \citenamefont {Addicoat}, \citenamefont {Gorelik},
  \citenamefont {Hambsch}, \citenamefont {Mundszinger}, \citenamefont {Park},
  \citenamefont {Lotsch} \emph {et~al.}}]{qi2020near}%
  \BibitemOpen
  \bibfield  {author} {\bibinfo {author} {\bibnamefont {Qi}, \bibfnamefont
  {H.}}, \bibinfo {author} {\bibnamefont {Sahabudeen}, \bibfnamefont {H.}},
  \bibinfo {author} {\bibnamefont {Liang}, \bibfnamefont {B.}}, \bibinfo
  {author} {\bibnamefont {Polo{\v{z}}ij}, \bibfnamefont {M.}}, \bibinfo
  {author} {\bibnamefont {Addicoat}, \bibfnamefont {M.~A.}}, \bibinfo {author}
  {\bibnamefont {Gorelik}, \bibfnamefont {T.~E.}}, \bibinfo {author}
  {\bibnamefont {Hambsch}, \bibfnamefont {M.}}, \bibinfo {author} {\bibnamefont
  {Mundszinger}, \bibfnamefont {M.}}, \bibinfo {author} {\bibnamefont {Park},
  \bibfnamefont {S.}}, \bibinfo {author} {\bibnamefont {Lotsch}, \bibfnamefont
  {B.~V.}},  \emph {et~al.},\ }\bibfield  {title} {\enquote {\bibinfo {title}
  {Near--atomic-scale observation of grain boundaries in a layer-stacked
  two-dimensional polymer},}\ }\href@noop {} {\bibfield  {journal} {\bibinfo
  {journal} {Science advances}\ }\textbf {\bibinfo {volume} {6}},\ \bibinfo
  {pages} {eabb5976} (\bibinfo {year} {2020})}\BibitemShut {NoStop}%
\bibitem [{\citenamefont {Rakowski}\ \emph {et~al.}(2021)\citenamefont
  {Rakowski}, \citenamefont {Munshi}, \citenamefont {Henderson}, \citenamefont
  {Cholia}, \citenamefont {Chan},\ and\ \citenamefont
  {Ophus}}]{rakoski2021database}%
  \BibitemOpen
  \bibfield  {author} {\bibinfo {author} {\bibnamefont {Rakowski},
  \bibfnamefont {A.}}, \bibinfo {author} {\bibnamefont {Munshi}, \bibfnamefont
  {J.}}, \bibinfo {author} {\bibnamefont {Henderson}, \bibfnamefont {M.}},
  \bibinfo {author} {\bibnamefont {Cholia}, \bibfnamefont {S.}}, \bibinfo
  {author} {\bibnamefont {Chan}, \bibfnamefont {M.}}, and\ \bibinfo {author}
  {\bibnamefont {Ophus}, \bibfnamefont {C.}},\ }\bibfield  {title} {\enquote
  {\bibinfo {title} {A complete pipeline for deep learning workflows in
  transmission electron microscopy},}\ }\href@noop {} {\bibfield  {journal}
  {\bibinfo  {journal} {manuscript in preparation}\ } (\bibinfo {year}
  {2021})}\BibitemShut {NoStop}%
\bibitem [{\citenamefont {Ramasse}(2017)}]{ramasse2017twenty}%
  \BibitemOpen
  \bibfield  {author} {\bibinfo {author} {\bibnamefont {Ramasse}, \bibfnamefont
  {Q.~M.}},\ }\bibfield  {title} {\enquote {\bibinfo {title} {Twenty years
  after: How “aberration correction in the {STEM}” truly placed a “a
  synchrotron in a microscope”},}\ }\href@noop {} {\bibfield  {journal}
  {\bibinfo  {journal} {Ultramicroscopy}\ }\textbf {\bibinfo {volume} {180}},\
  \bibinfo {pages} {41--51} (\bibinfo {year} {2017})}\BibitemShut {NoStop}%
\bibitem [{\citenamefont {Rangel~DaCosta}\ \emph {et~al.}(2021)\citenamefont
  {Rangel~DaCosta}, \citenamefont {Brown}, \citenamefont {Pelz}, \citenamefont
  {Rakowski}, \citenamefont {Barber}, \citenamefont {O’Donovan},
  \citenamefont {McBean}, \citenamefont {Jones}, \citenamefont {Ciston},
  \citenamefont {Scott} \emph {et~al.}}]{dacosta2021prismatic}%
  \BibitemOpen
  \bibfield  {author} {\bibinfo {author} {\bibnamefont {Rangel~DaCosta},
  \bibfnamefont {L.}}, \bibinfo {author} {\bibnamefont {Brown}, \bibfnamefont
  {H.~G.}}, \bibinfo {author} {\bibnamefont {Pelz}, \bibfnamefont {P.~M.}},
  \bibinfo {author} {\bibnamefont {Rakowski}, \bibfnamefont {A.}}, \bibinfo
  {author} {\bibnamefont {Barber}, \bibfnamefont {N.}}, \bibinfo {author}
  {\bibnamefont {O’Donovan}, \bibfnamefont {P.}}, \bibinfo {author}
  {\bibnamefont {McBean}, \bibfnamefont {P.}}, \bibinfo {author} {\bibnamefont
  {Jones}, \bibfnamefont {L.}}, \bibinfo {author} {\bibnamefont {Ciston},
  \bibfnamefont {J.}}, \bibinfo {author} {\bibnamefont {Scott}, \bibfnamefont
  {M.}},  \emph {et~al.},\ }\bibfield  {title} {\enquote {\bibinfo {title}
  {Prismatic 2.0-simulation software for scanning and high resolution
  transmission electron microscopy ({STEM} and {HRTEM})},}\ }\href@noop {}
  {\bibfield  {journal} {\bibinfo  {journal} {Micron}\ ,\ \bibinfo {pages}
  {103141}} (\bibinfo {year} {2021})}\BibitemShut {NoStop}%
\bibitem [{\citenamefont {Rauch}\ and\ \citenamefont
  {Dupuy}(2005)}]{rauch2005rapid}%
  \BibitemOpen
  \bibfield  {author} {\bibinfo {author} {\bibnamefont {Rauch}, \bibfnamefont
  {E.}}and\ \bibinfo {author} {\bibnamefont {Dupuy}, \bibfnamefont {L.}},\
  }\bibfield  {title} {\enquote {\bibinfo {title} {Rapid spot diffraction
  patterns idendification through template matching},}\ }\href@noop {}
  {\bibfield  {journal} {\bibinfo  {journal} {Archives of Metallurgy and
  Materials}\ }\textbf {\bibinfo {volume} {50}},\ \bibinfo {pages} {87--99}
  (\bibinfo {year} {2005})}\BibitemShut {NoStop}%
\bibitem [{\citenamefont {Rouviere}\ \emph {et~al.}(2013)\citenamefont
  {Rouviere}, \citenamefont {B{\'e}ch{\'e}}, \citenamefont {Martin},
  \citenamefont {Denneulin},\ and\ \citenamefont
  {Cooper}}]{rouviere2013improved}%
  \BibitemOpen
  \bibfield  {author} {\bibinfo {author} {\bibnamefont {Rouviere},
  \bibfnamefont {J.-L.}}, \bibinfo {author} {\bibnamefont {B{\'e}ch{\'e}},
  \bibfnamefont {A.}}, \bibinfo {author} {\bibnamefont {Martin}, \bibfnamefont
  {Y.}}, \bibinfo {author} {\bibnamefont {Denneulin}, \bibfnamefont {T.}}, and\
  \bibinfo {author} {\bibnamefont {Cooper}, \bibfnamefont {D.}},\ }\bibfield
  {title} {\enquote {\bibinfo {title} {Improved strain precision with high
  spatial resolution using nanobeam precession electron diffraction},}\
  }\href@noop {} {\bibfield  {journal} {\bibinfo  {journal} {Applied Physics
  Letters}\ }\textbf {\bibinfo {volume} {103}},\ \bibinfo {pages} {241913}
  (\bibinfo {year} {2013})}\BibitemShut {NoStop}%
\bibitem [{\citenamefont {Rowenhorst}\ \emph {et~al.}(2015)\citenamefont
  {Rowenhorst}, \citenamefont {Rollett}, \citenamefont {Rohrer}, \citenamefont
  {Groeber}, \citenamefont {Jackson}, \citenamefont {Konijnenberg},\ and\
  \citenamefont {De~Graef}}]{rowenhorst2015consistent}%
  \BibitemOpen
  \bibfield  {author} {\bibinfo {author} {\bibnamefont {Rowenhorst},
  \bibfnamefont {D.}}, \bibinfo {author} {\bibnamefont {Rollett}, \bibfnamefont
  {A.}}, \bibinfo {author} {\bibnamefont {Rohrer}, \bibfnamefont {G.}},
  \bibinfo {author} {\bibnamefont {Groeber}, \bibfnamefont {M.}}, \bibinfo
  {author} {\bibnamefont {Jackson}, \bibfnamefont {M.}}, \bibinfo {author}
  {\bibnamefont {Konijnenberg}, \bibfnamefont {P.~J.}}, and\ \bibinfo {author}
  {\bibnamefont {De~Graef}, \bibfnamefont {M.}},\ }\bibfield  {title} {\enquote
  {\bibinfo {title} {Consistent representations of and conversions between 3d
  rotations},}\ }\href@noop {} {\bibfield  {journal} {\bibinfo  {journal}
  {Modelling and Simulation in Materials Science and Engineering}\ }\textbf
  {\bibinfo {volume} {23}},\ \bibinfo {pages} {083501} (\bibinfo {year}
  {2015})}\BibitemShut {NoStop}%
\bibitem [{\citenamefont {Savitzky}\ \emph {et~al.}(2021)\citenamefont
  {Savitzky}, \citenamefont {Zeltmann}, \citenamefont {Hughes}, \citenamefont
  {Brown}, \citenamefont {Zhao}, \citenamefont {Pelz}, \citenamefont {Pekin},
  \citenamefont {Barnard}, \citenamefont {Donohue}, \citenamefont {DaCosta},
  \citenamefont {Kennedy}, \citenamefont {Xie}, \citenamefont {Janish},
  \citenamefont {Schneider}, \citenamefont {Herring}, \citenamefont {Gopal},
  \citenamefont {Anapolsky}, \citenamefont {Dhall}, \citenamefont {Bustillo},
  \citenamefont {Ercius}, \citenamefont {Scott}, \citenamefont {Ciston},
  \citenamefont {Minor},\ and\ \citenamefont {Ophus}}]{savitzky2021py4dstem}%
  \BibitemOpen
  \bibfield  {author} {\bibinfo {author} {\bibnamefont {Savitzky},
  \bibfnamefont {B.~H.}}, \bibinfo {author} {\bibnamefont {Zeltmann},
  \bibfnamefont {S.~E.}}, \bibinfo {author} {\bibnamefont {Hughes},
  \bibfnamefont {L.~A.}}, \bibinfo {author} {\bibnamefont {Brown},
  \bibfnamefont {H.~G.}}, \bibinfo {author} {\bibnamefont {Zhao}, \bibfnamefont
  {S.}}, \bibinfo {author} {\bibnamefont {Pelz}, \bibfnamefont {P.~M.}},
  \bibinfo {author} {\bibnamefont {Pekin}, \bibfnamefont {T.~C.}}, \bibinfo
  {author} {\bibnamefont {Barnard}, \bibfnamefont {E.~S.}}, \bibinfo {author}
  {\bibnamefont {Donohue}, \bibfnamefont {J.}}, \bibinfo {author} {\bibnamefont
  {DaCosta}, \bibfnamefont {L.~R.~D.}}, \bibinfo {author} {\bibnamefont
  {Kennedy}, \bibfnamefont {E.}}, \bibinfo {author} {\bibnamefont {Xie},
  \bibfnamefont {Y.}}, \bibinfo {author} {\bibnamefont {Janish}, \bibfnamefont
  {M.~T.}}, \bibinfo {author} {\bibnamefont {Schneider}, \bibfnamefont
  {M.~M.}}, \bibinfo {author} {\bibnamefont {Herring}, \bibfnamefont {P.}},
  \bibinfo {author} {\bibnamefont {Gopal}, \bibfnamefont {C.}}, \bibinfo
  {author} {\bibnamefont {Anapolsky}, \bibfnamefont {A.}}, \bibinfo {author}
  {\bibnamefont {Dhall}, \bibfnamefont {R.}}, \bibinfo {author} {\bibnamefont
  {Bustillo}, \bibfnamefont {K.~C.}}, \bibinfo {author} {\bibnamefont {Ercius},
  \bibfnamefont {P.}}, \bibinfo {author} {\bibnamefont {Scott}, \bibfnamefont
  {M.~C.}}, \bibinfo {author} {\bibnamefont {Ciston}, \bibfnamefont {J.}},
  \bibinfo {author} {\bibnamefont {Minor}, \bibfnamefont {A.~M.}}, and\
  \bibinfo {author} {\bibnamefont {Ophus}, \bibfnamefont {C.}},\ }\bibfield
  {title} {\enquote {\bibinfo {title} {{py4DSTEM}: A software package for
  four-dimensional scanning transmission electron microscopy data analysis},}\
  }\href@noop {} {\bibfield  {journal} {\bibinfo  {journal} {Microscopy and
  Microanalysis}\ }\textbf {\bibinfo {volume} {27}},\ \bibinfo {pages} {712}
  (\bibinfo {year} {2021})}\BibitemShut {NoStop}%
\bibitem [{\citenamefont {Shoemake}(1985)}]{shoemake1985animating}%
  \BibitemOpen
  \bibfield  {author} {\bibinfo {author} {\bibnamefont {Shoemake},
  \bibfnamefont {K.}},\ }\bibfield  {title} {\enquote {\bibinfo {title}
  {Animating rotation with quaternion curves},}\ }in\ \href@noop {} {\emph
  {\bibinfo {booktitle} {Proceedings of the 12th annual conference on Computer
  graphics and interactive techniques}}}\ (\bibinfo {year} {1985})\ pp.\
  \bibinfo {pages} {245--254}\BibitemShut {NoStop}%
\bibitem [{\citenamefont {Spence}(1993)}]{spence1993accurate}%
  \BibitemOpen
  \bibfield  {author} {\bibinfo {author} {\bibnamefont {Spence}, \bibfnamefont
  {J.}},\ }\bibfield  {title} {\enquote {\bibinfo {title} {On the accurate
  measurement of structure-factor amplitudes and phases by electron
  diffraction},}\ }\href@noop {} {\bibfield  {journal} {\bibinfo  {journal}
  {Acta Crystallographica Section A: Foundations of Crystallography}\ }\textbf
  {\bibinfo {volume} {49}},\ \bibinfo {pages} {231--260} (\bibinfo {year}
  {1993})}\BibitemShut {NoStop}%
\bibitem [{\citenamefont {Spurgeon}\ \emph {et~al.}(2021)\citenamefont
  {Spurgeon}, \citenamefont {Ophus}, \citenamefont {Jones}, \citenamefont
  {Petford-Long}, \citenamefont {Kalinin}, \citenamefont {Olszta},
  \citenamefont {Dunin-Borkowski}, \citenamefont {Salmon}, \citenamefont
  {Hattar}, \citenamefont {Yang} \emph {et~al.}}]{spurgeon2021towards}%
  \BibitemOpen
  \bibfield  {author} {\bibinfo {author} {\bibnamefont {Spurgeon},
  \bibfnamefont {S.~R.}}, \bibinfo {author} {\bibnamefont {Ophus},
  \bibfnamefont {C.}}, \bibinfo {author} {\bibnamefont {Jones}, \bibfnamefont
  {L.}}, \bibinfo {author} {\bibnamefont {Petford-Long}, \bibfnamefont {A.}},
  \bibinfo {author} {\bibnamefont {Kalinin}, \bibfnamefont {S.~V.}}, \bibinfo
  {author} {\bibnamefont {Olszta}, \bibfnamefont {M.~J.}}, \bibinfo {author}
  {\bibnamefont {Dunin-Borkowski}, \bibfnamefont {R.~E.}}, \bibinfo {author}
  {\bibnamefont {Salmon}, \bibfnamefont {N.}}, \bibinfo {author} {\bibnamefont
  {Hattar}, \bibfnamefont {K.}}, \bibinfo {author} {\bibnamefont {Yang},
  \bibfnamefont {W.-C.~D.}},  \emph {et~al.},\ }\bibfield  {title} {\enquote
  {\bibinfo {title} {Towards data-driven next-generation transmission electron
  microscopy},}\ }\href@noop {} {\bibfield  {journal} {\bibinfo  {journal}
  {Nature materials}\ }\textbf {\bibinfo {volume} {20}},\ \bibinfo {pages}
  {274--279} (\bibinfo {year} {2021})}\BibitemShut {NoStop}%
\bibitem [{\citenamefont {Tang}, \citenamefont {Carter},\ and\ \citenamefont
  {Cannon}(2006)}]{tang2006diffuse}%
  \BibitemOpen
  \bibfield  {author} {\bibinfo {author} {\bibnamefont {Tang}, \bibfnamefont
  {M.}}, \bibinfo {author} {\bibnamefont {Carter}, \bibfnamefont {W.~C.}}, and\
  \bibinfo {author} {\bibnamefont {Cannon}, \bibfnamefont {R.~M.}},\ }\bibfield
   {title} {\enquote {\bibinfo {title} {Diffuse interface model for structural
  transitions of grain boundaries},}\ }\href@noop {} {\bibfield  {journal}
  {\bibinfo  {journal} {Physical Review B}\ }\textbf {\bibinfo {volume} {73}},\
  \bibinfo {pages} {024102} (\bibinfo {year} {2006})}\BibitemShut {NoStop}%
\bibitem [{\citenamefont {Tao}\ \emph {et~al.}(2009)\citenamefont {Tao},
  \citenamefont {Niebieskikwiat}, \citenamefont {Varela}, \citenamefont {Luo},
  \citenamefont {Schofield}, \citenamefont {Zhu}, \citenamefont {Salamon},
  \citenamefont {Zuo}, \citenamefont {Pantelides},\ and\ \citenamefont
  {Pennycook}}]{tao2009direct}%
  \BibitemOpen
  \bibfield  {author} {\bibinfo {author} {\bibnamefont {Tao}, \bibfnamefont
  {J.}}, \bibinfo {author} {\bibnamefont {Niebieskikwiat}, \bibfnamefont {D.}},
  \bibinfo {author} {\bibnamefont {Varela}, \bibfnamefont {M.}}, \bibinfo
  {author} {\bibnamefont {Luo}, \bibfnamefont {W.}}, \bibinfo {author}
  {\bibnamefont {Schofield}, \bibfnamefont {M.}}, \bibinfo {author}
  {\bibnamefont {Zhu}, \bibfnamefont {Y.}}, \bibinfo {author} {\bibnamefont
  {Salamon}, \bibfnamefont {M.~B.}}, \bibinfo {author} {\bibnamefont {Zuo},
  \bibfnamefont {J.-M.}}, \bibinfo {author} {\bibnamefont {Pantelides},
  \bibfnamefont {S.~T.}}, and\ \bibinfo {author} {\bibnamefont {Pennycook},
  \bibfnamefont {S.~J.}},\ }\bibfield  {title} {\enquote {\bibinfo {title}
  {Direct imaging of nanoscale phase separation in la 0.55 ca 0.45 mno 3:
  relationship to colossal magnetoresistance},}\ }\href@noop {} {\bibfield
  {journal} {\bibinfo  {journal} {Physical Review Letters}\ }\textbf {\bibinfo
  {volume} {103}},\ \bibinfo {pages} {097202} (\bibinfo {year}
  {2009})}\BibitemShut {NoStop}%
\bibitem [{\citenamefont {Thompson}(2000)}]{thompson2000structure}%
  \BibitemOpen
  \bibfield  {author} {\bibinfo {author} {\bibnamefont {Thompson},
  \bibfnamefont {C.~V.}},\ }\bibfield  {title} {\enquote {\bibinfo {title}
  {Structure evolution during processing of polycrystalline films},}\
  }\href@noop {} {\bibfield  {journal} {\bibinfo  {journal} {Annual review of
  materials science}\ }\textbf {\bibinfo {volume} {30}},\ \bibinfo {pages}
  {159--190} (\bibinfo {year} {2000})}\BibitemShut {NoStop}%
\bibitem [{\citenamefont {Thompson}\ and\ \citenamefont
  {Carel}(1995)}]{thompson1995texture}%
  \BibitemOpen
  \bibfield  {author} {\bibinfo {author} {\bibnamefont {Thompson},
  \bibfnamefont {C.~V.}}and\ \bibinfo {author} {\bibnamefont {Carel},
  \bibfnamefont {R.}},\ }\bibfield  {title} {\enquote {\bibinfo {title}
  {Texture development in polycrystalline thin films},}\ }\href@noop {}
  {\bibfield  {journal} {\bibinfo  {journal} {Materials Science and
  Engineering: B}\ }\textbf {\bibinfo {volume} {32}},\ \bibinfo {pages}
  {211--219} (\bibinfo {year} {1995})}\BibitemShut {NoStop}%
\bibitem [{\citenamefont {Wang}\ \emph {et~al.}(2011)\citenamefont {Wang},
  \citenamefont {Wang}, \citenamefont {Sun}, \citenamefont {Zhang},
  \citenamefont {Chen}, \citenamefont {Wang}, \citenamefont {Shen},
  \citenamefont {Han}, \citenamefont {Lu},\ and\ \citenamefont
  {Chen}}]{wang2011}%
  \BibitemOpen
  \bibfield  {author} {\bibinfo {author} {\bibnamefont {Wang}, \bibfnamefont
  {Y.}}, \bibinfo {author} {\bibnamefont {Wang}, \bibfnamefont {Q.}}, \bibinfo
  {author} {\bibnamefont {Sun}, \bibfnamefont {H.}}, \bibinfo {author}
  {\bibnamefont {Zhang}, \bibfnamefont {W.}}, \bibinfo {author} {\bibnamefont
  {Chen}, \bibfnamefont {G.}}, \bibinfo {author} {\bibnamefont {Wang},
  \bibfnamefont {Y.}}, \bibinfo {author} {\bibnamefont {Shen}, \bibfnamefont
  {X.}}, \bibinfo {author} {\bibnamefont {Han}, \bibfnamefont {Y.}}, \bibinfo
  {author} {\bibnamefont {Lu}, \bibfnamefont {X.}}, and\ \bibinfo {author}
  {\bibnamefont {Chen}, \bibfnamefont {H.}},\ }\bibfield  {title} {\enquote
  {\bibinfo {title} {{Chiral Transformation: From Single Nanowire to Double
  Helix}},}\ }\href@noop {} {\bibfield  {journal} {\bibinfo  {journal} {Journal
  of the American Chemical Society}\ }\textbf {\bibinfo {volume} {133}},\
  \bibinfo {pages} {20060--20063} (\bibinfo {year} {2011})}\BibitemShut
  {NoStop}%
\bibitem [{\citenamefont {Wright}, \citenamefont {Nowell},\ and\ \citenamefont
  {Field}(2011)}]{wright2011review}%
  \BibitemOpen
  \bibfield  {author} {\bibinfo {author} {\bibnamefont {Wright}, \bibfnamefont
  {S.~I.}}, \bibinfo {author} {\bibnamefont {Nowell}, \bibfnamefont {M.~M.}},
  and\ \bibinfo {author} {\bibnamefont {Field}, \bibfnamefont {D.~P.}},\
  }\bibfield  {title} {\enquote {\bibinfo {title} {A review of strain analysis
  using electron backscatter diffraction},}\ }\href@noop {} {\bibfield
  {journal} {\bibinfo  {journal} {Microscopy and microanalysis}\ }\textbf
  {\bibinfo {volume} {17}},\ \bibinfo {pages} {316--329} (\bibinfo {year}
  {2011})}\BibitemShut {NoStop}%
\bibitem [{\citenamefont {Wright}\ \emph {et~al.}(2015)\citenamefont {Wright},
  \citenamefont {Nowell}, \citenamefont {Lindeman}, \citenamefont {Camus},
  \citenamefont {De~Graef},\ and\ \citenamefont
  {Jackson}}]{wright2015introduction}%
  \BibitemOpen
  \bibfield  {author} {\bibinfo {author} {\bibnamefont {Wright}, \bibfnamefont
  {S.~I.}}, \bibinfo {author} {\bibnamefont {Nowell}, \bibfnamefont {M.~M.}},
  \bibinfo {author} {\bibnamefont {Lindeman}, \bibfnamefont {S.~P.}}, \bibinfo
  {author} {\bibnamefont {Camus}, \bibfnamefont {P.~P.}}, \bibinfo {author}
  {\bibnamefont {De~Graef}, \bibfnamefont {M.}}, and\ \bibinfo {author}
  {\bibnamefont {Jackson}, \bibfnamefont {M.~A.}},\ }\bibfield  {title}
  {\enquote {\bibinfo {title} {Introduction and comparison of new ebsd
  post-processing methodologies},}\ }\href@noop {} {\bibfield  {journal}
  {\bibinfo  {journal} {Ultramicroscopy}\ }\textbf {\bibinfo {volume} {159}},\
  \bibinfo {pages} {81--94} (\bibinfo {year} {2015})}\BibitemShut {NoStop}%
\bibitem [{\citenamefont {Wu}\ \emph {et~al.}(2021)\citenamefont {Wu},
  \citenamefont {Harreiss}, \citenamefont {Ophus},\ and\ \citenamefont
  {Spiecker}}]{wu2021seeing}%
  \BibitemOpen
  \bibfield  {author} {\bibinfo {author} {\bibnamefont {Wu}, \bibfnamefont
  {M.}}, \bibinfo {author} {\bibnamefont {Harreiss}, \bibfnamefont {C.}},
  \bibinfo {author} {\bibnamefont {Ophus}, \bibfnamefont {C.}}, and\ \bibinfo
  {author} {\bibnamefont {Spiecker}, \bibfnamefont {E.}},\ }\bibfield  {title}
  {\enquote {\bibinfo {title} {Seeing structural evolution of organic molecular
  nano-crystallites using 4d scanning confocal electron diffraction},}\
  }\href@noop {} {\bibfield  {journal} {\bibinfo  {journal} {arXiv preprint
  arXiv:2110.02373}\ } (\bibinfo {year} {2021})}\BibitemShut {NoStop}%
\bibitem [{\citenamefont {Yuan}\ \emph {et~al.}(2021)\citenamefont {Yuan},
  \citenamefont {Zhang}, \citenamefont {He},\ and\ \citenamefont
  {Zuo}}]{yuan2021training}%
  \BibitemOpen
  \bibfield  {author} {\bibinfo {author} {\bibnamefont {Yuan}, \bibfnamefont
  {R.}}, \bibinfo {author} {\bibnamefont {Zhang}, \bibfnamefont {J.}}, \bibinfo
  {author} {\bibnamefont {He}, \bibfnamefont {L.}}, and\ \bibinfo {author}
  {\bibnamefont {Zuo}, \bibfnamefont {J.-M.}},\ }\bibfield  {title} {\enquote
  {\bibinfo {title} {Training artificial neural networks for precision
  orientation and strain mapping using 4d electron diffraction datasets},}\
  }\href@noop {} {\bibfield  {journal} {\bibinfo  {journal} {Ultramicroscopy}\
  ,\ \bibinfo {pages} {113256}} (\bibinfo {year} {2021})}\BibitemShut {NoStop}%
\bibitem [{\citenamefont {Zaefferer}\ and\ \citenamefont
  {Schwarzer}(1994)}]{zaefferer1994line}%
  \BibitemOpen
  \bibfield  {author} {\bibinfo {author} {\bibnamefont {Zaefferer},
  \bibfnamefont {S.}}and\ \bibinfo {author} {\bibnamefont {Schwarzer},
  \bibfnamefont {R.~A.}},\ }\bibfield  {title} {\enquote {\bibinfo {title}
  {On-line interpretation of spot and kikuchi patterns},}\ }in\ \href@noop {}
  {\emph {\bibinfo {booktitle} {Materials Science Forum}}},\ Vol.\ \bibinfo
  {volume} {157}\ (\bibinfo {organization} {Trans Tech Publ},\ \bibinfo {year}
  {1994})\ pp.\ \bibinfo {pages} {247--250}\BibitemShut {NoStop}%
\bibitem [{\citenamefont {Zeltmann}\ \emph {et~al.}(2020)\citenamefont
  {Zeltmann}, \citenamefont {Muller}, \citenamefont {Bustillo}, \citenamefont
  {Savitsky}, \citenamefont {Hughes}, \citenamefont {Minor},\ and\
  \citenamefont {Ophus}}]{zeltmann2020}%
  \BibitemOpen
  \bibfield  {author} {\bibinfo {author} {\bibnamefont {Zeltmann},
  \bibfnamefont {S.~E.}}, \bibinfo {author} {\bibnamefont {Muller},
  \bibfnamefont {A.}}, \bibinfo {author} {\bibnamefont {Bustillo},
  \bibfnamefont {K.~C.}}, \bibinfo {author} {\bibnamefont {Savitsky},
  \bibfnamefont {B.}}, \bibinfo {author} {\bibnamefont {Hughes}, \bibfnamefont
  {L.}}, \bibinfo {author} {\bibnamefont {Minor}, \bibfnamefont {A.~M.}}, and\
  \bibinfo {author} {\bibnamefont {Ophus}, \bibfnamefont {C.}},\ }\bibfield
  {title} {\enquote {\bibinfo {title} {{Patterned probes for high precision
  4D-STEM Bragg measurements}},}\ }\href@noop {} {\bibfield  {journal}
  {\bibinfo  {journal} {Ultramicroscopy}\ }\textbf {\bibinfo {volume} {209}},\
  \bibinfo {pages} {112890} (\bibinfo {year} {2020})}\BibitemShut {NoStop}%
\bibitem [{\citenamefont {Zuo}\ and\ \citenamefont
  {Zhu}(2021)}]{zuo2021strategies}%
  \BibitemOpen
  \bibfield  {author} {\bibinfo {author} {\bibnamefont {Zuo}, \bibfnamefont
  {J.-M.}}and\ \bibinfo {author} {\bibnamefont {Zhu}, \bibfnamefont {X.}},\
  }\bibfield  {title} {\enquote {\bibinfo {title} {Strategies for fast and
  reliable {4D-STEM} orientation and phase mapping of nanomaterials and
  devices},}\ }\href@noop {} {\bibfield  {journal} {\bibinfo  {journal}
  {Microscopy and Microanalysis}\ }\textbf {\bibinfo {volume} {27}},\ \bibinfo
  {pages} {762--763} (\bibinfo {year} {2021})}\BibitemShut {NoStop}%
\end{thebibliography}%

\end{document}